\documentclass[useAMS,usenatbib]{mn2e}
\usepackage{epsf}
\usepackage{amssymb}
\usepackage[usenames]{color}
\usepackage{graphicx}	% Including figure files
\usepackage{amsmath}	% Advanced maths commands
\usepackage{savesym}
\usepackage{natbib}
\usepackage{epstopdf}
\usepackage{mathrsfs,dsfont}
\usepackage{multirow}
\usepackage{subfigure}
\usepackage{tikz}

\usepackage{color}
\definecolor{grey}{rgb}{0.4,0.6,0.6}
\definecolor{darkgreen}{rgb}{0.,0.7,0.}

\def\msun{{M_\odot}}

\def\kms{{\rm \,km\,s^{-1}}}

\def\Gyr{{\rm \,Gyr}}

\def\kpc{{\rm \,kpc}}

\def\msun{{M_\odot}}

\newcommand{\be}{\begin{equation}}
\newcommand{\ee}{\end{equation}}

\newcommand{\xmm}{\textit{XMM-Newton} }
\newcommand{\chandra}{\textit{Chandra} }

\newcommand{\dg}{$^{\circ}$}

\title[NGC 4839 group merging with Coma ]{Close-up view of an ongoing merger between the NGC~4839 group and the Coma cluster - a post-merger scenario
}
\author[Lyskova et al.]{N.~Lyskova$^{1,2,3,4}$\thanks{E-mail: natalya.lyskova@gmail.com}, 
  E.~Churazov$^{4,2}$\thanks{E-mail: churazov@mpa-garching.mpg.de}, C.~Zhang$^{4}$, W. Forman$^{5}$, C. Jones$^{5}$, \newauthor
  K.~Dolag$^{6,4}$, E.~Roediger$^{7}$, A.~Sheardown$^{7}$ \\
$^1$ National Research University Higher School of Economics, Myasnitskaya str. 
  20, Moscow 101000, Russia\\
$^2$ Space Research Institute (IKI), Profsoyuznaya 84/32, Moscow 117997, 
Russia\\
$^3$ ASC of P.N.Lebedev Physical Institute, Leninskiy prospect 53, Moscow 119991, Russia \\  
$^4$ Max-Planck-Institut f\"ur Astrophysik, Karl-Schwarzschild-Strasse 1, 85741
Garching, Germany\\
$^5$ Harvard-Smithsonian Center for Astrophysics, 60 Garden Street, Cambridge, MA 02138, USA\\
$^6$ University Observatory Munich, Scheinerstr. 1, 81679 Munich, Germany\\
$^7$ E.A. Milne Centre for Astrophysics, School of Mathematics and Physical Sciences, University of Hull, Hull, HU6 7RX, United Kingdom \\
}

\begin{document}
%\date{Accepted ????????????;  Received ????????????; in original form ????????????}

\pagerange{\pageref{firstpage}--\pageref{lastpage}}
\pubyear{2018}

\maketitle

\label{firstpage}
\begin{abstract}

We study a merger of the NGC~4839 group with the Coma cluster using X-ray observations from the \textit{XMM-Newton} and \textit{Chandra} telescopes. X-ray data show two prominent features: (i) a long ($\sim$600 kpc in projection) and bent tail of cool gas trailing (towards south-west) the optical center of NGC~4839, and ii) a `sheath' region of enhanced X-ray surface brightness enveloping the group, which is due to hotter gas. While at  first glance the X-ray images suggest that we are witnessing the first infall of NGC~4839 into the Coma cluster core, we argue that a post-merger scenario provides a better explanation of the observed features and illustrate this with a series of numerical simulations. In this scenario, the tail is formed when the group, initially moving to the south-west, reverses its radial velocity after crossing the \mbox{apocenter}, the ram pressure ceases and the ram-pressure-displaced gas falls back toward the center of the group and overshoots it. Shortly after the \mbox{apocenter} passage, the optical galaxy, dark matter and gaseous core move in a north-east direction, while the displaced gas continues moving to the south-west. The `sheath' is explained as being due to interaction of the re-infalling group with its own tail of stripped gas mixed with the Coma gas. In this scenario, the shock, driven by the group before reaching the \mbox{apocenter}, has already detached from the group and would be located close to the famous relic to the south-west of the Coma cluster.

\end{abstract}


%
%________________________________________________________________

\sloppypar

\section{Introduction}

According to the current cosmological concordance model, galaxy clusters form  hierarchically through a combination of rare major merger events and  more gentle continuous accretion of smaller groups of galaxies throughout cosmic time along  filamentary structures \citep[e.g.][]{2012ARA&A..50..353K,2014PhyU...57..317V}. % \citep[][and many others]{2006Natur.440.1137S}.
The latter actually represents the dominant channel of galaxy cluster growth in mass and in the number of member galaxies \citep[see, e.g.,][]{2009MNRAS.399..497D, 2009ApJ...690.1292B, 2010ApJ...719..229G}.
The best sites for studying structure formation processes are the galaxy cluster outskirts, which represent the transition region between the virialized intra-cluster medium (ICM) approximately in hydrostatic equilibrium, and the infalling material from the surroundings.
Mergers and accretion events often leave imprints on the distribution of the hot ICM such as shocks, cold fronts, and ram pressure stripped tails which could be observed with X-ray observations   \citep[e.g.,][among others]{2006ApJ...637L..81S, 2007PhR...443....1M, 2010ApJ...708..946S, 2017ApJ...835...19S}. 
In particular, the appearance of extended tails of ram pressure stripped gas provides valuable information for recovering the 3D geometry of a merger event, the orbital stage of a subhalo (early infall, pre-/post-pericenter passage), and even ICM plasma properties such as viscosity, magnetic fields, thermal conduction \citep[][among others]{2015ApJ...806..103R,2015ApJ...806..104R}. 
Structure formation processes in galaxy clusters can be studied with radio observations of radio relics, which are believed to trace large-scale shock waves  generated by a merger or  accretion, e.g.,  \cite{1998A&A...332..395E, 2009A&A...505..991V, 2018ApJ...865...24D}.

The Coma cluster of galaxies (Abell 1656), the third brightest X-ray cluster,  is one of the nearest and best-studied galaxy clusters. Optical and X-ray observations have revealed a wealth of substructure in the Coma cluster \citep[][among others]{Colless.Dunn.1996, 1997ApJ...474L...7V, 2001A&A...365L..60B,  Adami.et.al.2005, Neumann.et.al.2003,2013ApJ...766..107A}.  One of the most prominent  is the group of galaxies associated with the elliptical galaxy NGC~4839. It lies in the cluster outskirts ($\sim 1$ Mpc in projection) south-west of the cluster center.
The X-ray image (see Fig.~\ref{fig:all}) exhibits an edge-like structure at the head of the group and an elongated tail of ram pressure stripped gas toward
the south-west, i.e. opposite to the direction to the Coma center. 
The NGC~4839 group appears to be merging with the Coma cluster core, and the tail direction
is approximately aligned with a filament connecting the Coma cluster
with Abell~1367, which, in turn, is part of the `Great Wall' \citep[e.g., ][]{1989Sci...246..897G,Neumann.et.al.2001,Brown.Rudnick.2011}. It has long been debated  whether observations are consistent with a simple radial infall or imply a tangential orbit \citep[][]{{1996A&A...311...95B}}, and whether we observe the first passage \citep[][among others]{Colless.Dunn.1996, Neumann.et.al.2001, Akamatsu.et.al.2013} or the group has already passed the Coma cluster center \citep[][]{Burns.et.al.1994}.    

Based on galaxy redshifts, \cite{Colless.Dunn.1996} constrained the 3D geometry of the Coma-NGC~4839 merger using 
a simple dynamical two-body model, in which the two clusters were considered as point masses following a linear orbit under their mutual gravity. They
 estimated that the angle between merging objects and the observer is most likely to be $\alpha = 74$\dg$^{+5}_{-10}$, i.e. the merger happens almost in the plane of the sky, the true 3D separation is $0.8\pm 0.1$ $h^{-1}$ Mpc and the infall velocity is $1700^{+350}_{-500}$ km s$^{-1}$. \cite{Colless.Dunn.1996} argued that the NGC~4839 group is just beginning to penetrate the Coma cluster.

Coma, as a typical merging cluster, also hosts a radio halo and a relic. The Coma radio relic is located  $\sim 2.1$ Mpc in projection from the cluster center, beyond the NGC~4839 group, but in the same south-west direction. Based on radio and optical
observations, \cite{Brown.Rudnick.2011} suggested that the Coma radio relic is due to an infall shock, caused  by  the infall  of a `wall' of galaxies possibly associated with NGC~4839 into the Coma cluster. 
However, large radio relics (radio gischt), in general, are believed to be associated with outgoing merger shocks \citep[e.g.,][among others]{2009A&A...494..429B, 2010Sci...330..347V}.  
 Analyses of \xmm  \citep{Ogrean.Bruggen.2013} and \textit{Suzaku} observations  \citep{Akamatsu.et.al.2013}  revealed a tentative temperature discontinuity across the Coma relic, which has been interpreted as a shock front with a Mach number of $M \sim 2$. 
%supporting the idea of \cite{Brown.Rudnick.2011} that the relic traces the infall shock.
Moreover, a tentative detection of a pressure jump at the position of the radio relic has been reported \citep{Erler.et.al.2015} based on the thermal Sunyaev-Zel'dovich effect data  extracted from the first public all-sky data release of \textit{Planck}. 

Based on a weak gravitational lensing survey performed with the Subaru/Suprime-Cam, \cite{2014PASJ...66...99O} detected a subhalo associated with the NGC~4839 group and measured its mass to be $\sim 10^{13} M_{\odot}$, assuming an NFW profile with a truncation  radius of $\sim 100$~kpc.
Curiously, the X-ray peak derived from the \textit{XMM-Newton} image is spatially coincident with the NGC~4839 elliptical galaxy, but both are shifted  $\simeq 1' \simeq 30 $ kpc towards the west from the weak-lensing mass center (see Fig. 2(b) in \citealt{Sasaki.et.al.2016}).

Here, we analyse available \xmm and \chandra observations. To constrain the  3D geometry and a trajectory of the infalling group, we compare the  X-ray maps with SPH simulations. 
In Sections~\ref{sec:coma} and \ref{sec:ngc4839} we describe general X-ray properties of the Coma cluster and the NGC~4839 group, respectively. The simulation setup is outlined in   Section~\ref{sec:setup}, followed by a discussion of  pre-merger and post-merger scenarios in Section~\ref{sec:results}. We discuss the mass and gas temperature of the NGC~4839 group in Section~\ref{sec:discuss}  and summarize our findings in Section~\ref{sec:conc}.

All our results are scaled to a flat $\Lambda$CDM cosmology with $\Omega_m=0.3$, $\Omega_{\Lambda}=0.7$, and a Hubble constant $H_0 = 70$ km s$^{-1}$ Mpc$^{-1}$, which implies a linear scale of 27.98 kpc arcmin$^{-1}$ at  Coma's redshift $z = 0.0231$ (NED\footnote{https://ned.ipac.caltech.edu} database). For the Coma cluster, we assume 
$r_{500} \simeq (47 \pm 1)$ arcmin $\simeq (1.315 \pm 0.028)$ Mpc and $r_{200} \simeq 1.5r_{500 }$ as in  \citet{Planck}, where $r_{500}$ and $r_{200}$ are the radii within which the mean density of the cluster is 500 and 200 times the critical density of the universe, respectively.

\section{Coma}
\label{sec:coma}

\subsection{X-ray surface brightness}
\label{subsec:beta}

\begin{figure*}%[b]
%\vspace*{-1.0 cm}
\vspace*{0.5 cm}
\centerline{
 \includegraphics[width=0.35\textwidth,clip=t,angle=0.,scale=0.98]{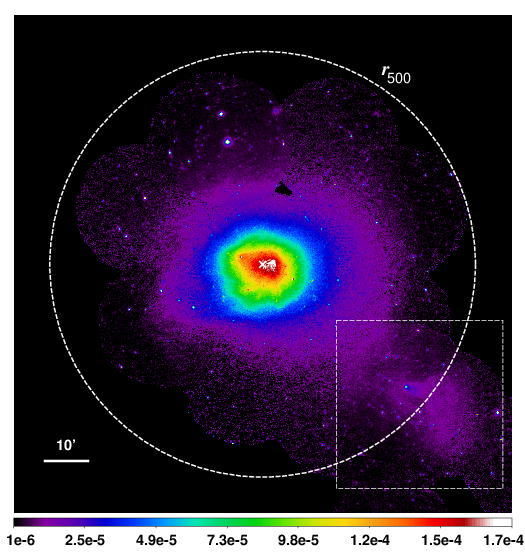}
 \includegraphics[width=0.35\textwidth,clip=t,angle=0.,scale=0.98]{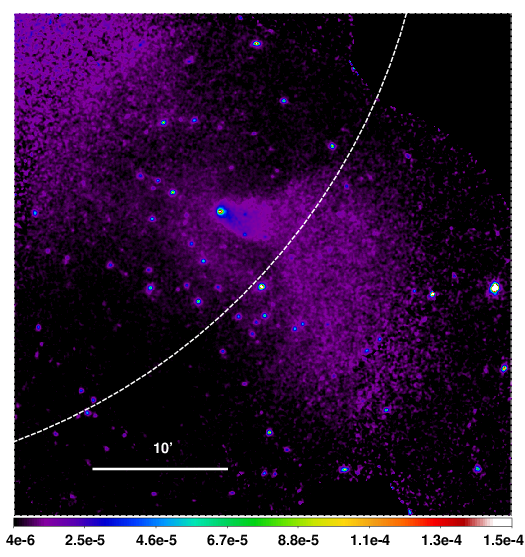}
}
%\vspace*{2.0 cm}
\centerline{
 \includegraphics[width=0.35\textwidth,clip=t,angle=0.,scale=0.98]{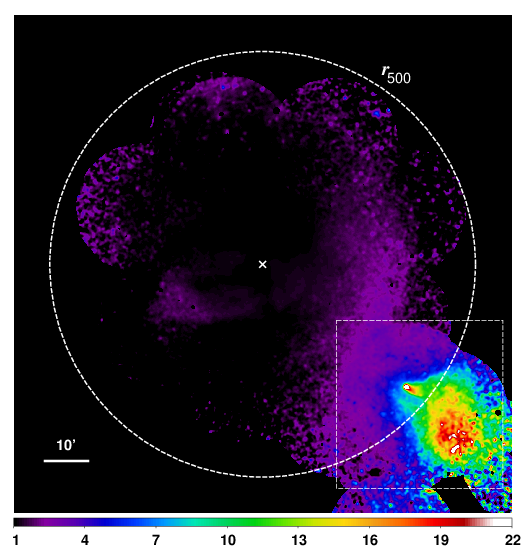}
 \includegraphics[width=0.35\textwidth,clip=t,angle=0.,scale=0.98]{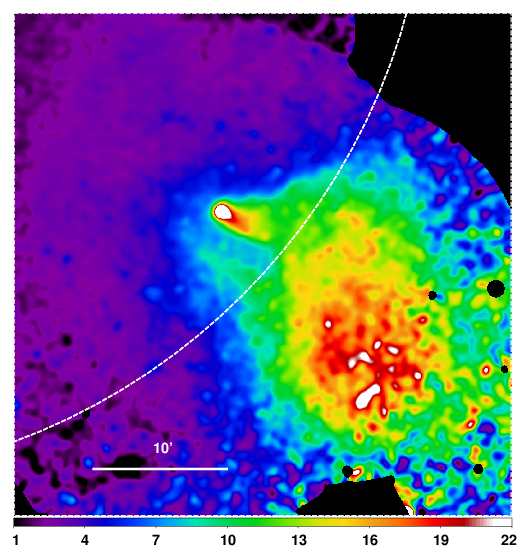}
}
%\vspace*{2.0 cm}
\centerline{
 \includegraphics[width=0.35\textwidth,clip=t,angle=0.,scale=0.98]{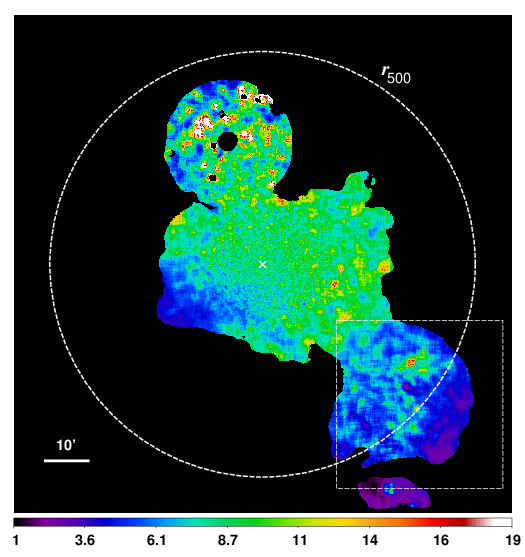}
 \includegraphics[width=0.35\textwidth,clip=t,angle=0.,scale=0.98]{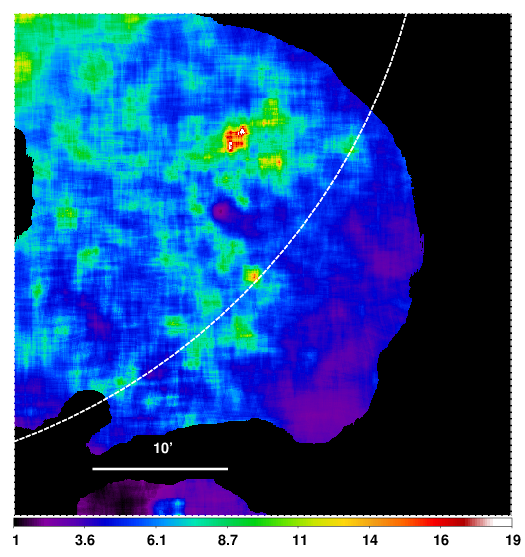}
}
\caption{\textit{Upper row:} \textit{XMM-Newton} image of the Coma Cluster (left panel) and the NGC~4839 group (right panel) in the 0.5-2.5 keV energy band.
\textit{Middle row:} Surface brightness image divided by the best fit $\beta$-model.
\textit{Bottom row:} Projected temperature map. The white dashed circle marks the radius of $r_{500}$ = 47~arcmin (\citealt{Planck}). The white scale bar indicates 10 arcmin.}
\label{fig:all}
\end{figure*}

For our analysis, we used publicly available \xmm  data, namely, the data from the
EPIC/MOS (European photon imaging camera/metal oxide semiconductor) detector.
The data were prepared by removing background flares using the light curve of
the detected events above 10~keV and renormalizing the `blank
fields' background to match the observed count rate in the $11 - 12$ keV band \citep[][]{2003ApJ...590..225C}.
The resulting background-subtracted, exposure- and vignetting-corrected \xmm image of the Coma cluster in the 0.5-2.5 keV energy band
is presented in Fig.~\ref{fig:all}  (upper row). 
The zoom-in view of the NGC~4839 group is shown in the right panel. 

%The total exposure of all used observations is XXX ks.

\begin{figure}
	\includegraphics[width=\columnwidth]{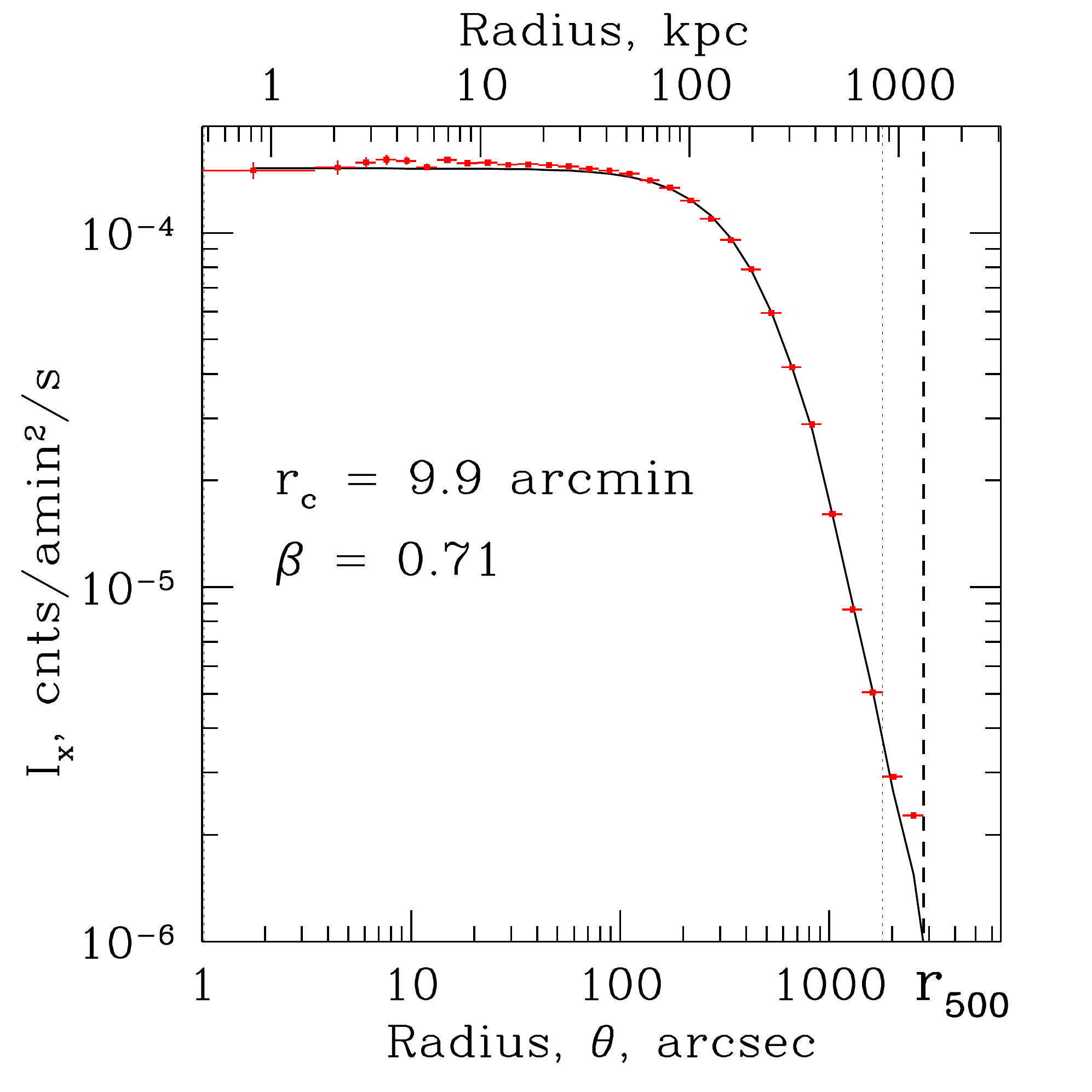}
    \caption{Radial X-ray surface brightness profile of the Coma cluster based on the \xmm data. The solid line shows the best-fit $\beta$-model with  core radius $r_c = 9.9$ arcmin $\simeq 277$ kpc and $\beta = 0.71 $. The sector containing NGC~4839 was excluded. The dashed vertical line marks $r_{500}$ and the dotted line shows the outer bound of the radial range used for fitting.}
    \label{fig:beta}
\end{figure}

For the surface brightness analysis, we extracted X-ray counts from within  circular annuli, with their center at (RA, DEC) = (194.94458, 27.95) =  (12:59:46.699, +27:57:00.00). The sector containing the NGC~4839 group was excluded. The azimuthally averaged surface brightness profile (Fig.~\ref{fig:beta}) 
is reasonably well fit with the one-dimensional $\beta$-model \citep{beta_model}:

\be
I(R) = I_o \left( 1 + \frac{R^2}{r_c^2} \right)^{1/2-3\beta}
\label{eq:beta}
\ee
with the core radius $r_c = 9.9$ arcmin $\simeq 277$ kpc and $\beta = 0.71 $. Here, $R$ is the projected distance from the cluster center and $I_0$ is the surface brightness at the center.  The best-fit $\beta$ surface brightness profile is shown in Fig.~\ref{fig:beta} by a solid black line. This model is used in the subsequent analyses. To remove the global cluster emission and highlight the surface brightness deviations from the smooth underlying model, we divide   the initial X-ray image  by the best-fitting spherically symmetric $\beta$-model.
The result is shown in the middle row of Fig.~\ref{fig:all}. In addition to the extended X-ray emission associated with the NGC~4839 group, there is a prominent X-ray excess (relative to the best-fit $\beta$-model) in the west, elongated in the north-south direction, and a filamentary shaped excess (though not very prominent in this image) to the east of the Coma centre, elongated  in  the  east-west  direction \citep{1997ApJ...474L...7V}. These substructures are also described in \cite{Neumann.et.al.2003}.

\subsection{Temperature map}
\label{sec:temp}

\begin{figure}
\centering
\includegraphics[width=0.4\textwidth]{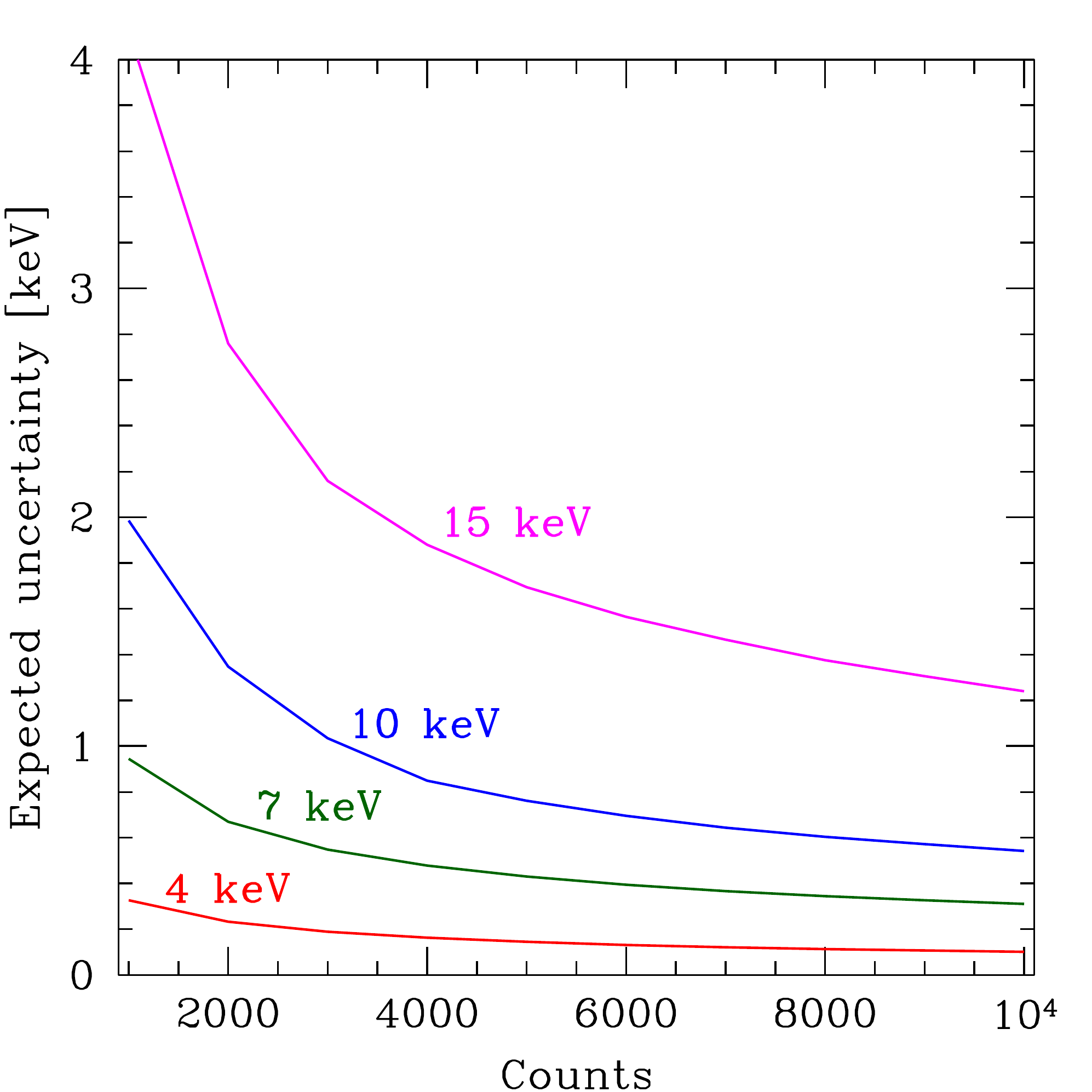}
\caption{Pure statistical uncertainties (arising from the photon counts) of the temperature measurements for different plasma temperatures (4, 7, 10, and 15 keV) plotted against source photon counts. In the projected temperature map shown in the bottom row of Fig.~\ref{fig:all}, each value of the temperature was calculated using regions with $\sim 3000$ photon counts.} 
\label{fig:tmap_uncertainty}
\end{figure}

Projected temperature maps can be generated by extracting spectra from regions with a sufficient number of photon counts 
and then  approximating  them  with  a  model  of  optically  thin plasma. In lieu of the direct fitting procedure, we used a technique described in \cite{Churazov.et.al.1996, Churazov.et.al.2016} to construct a projected temperature map of the Coma cluster. The method is computationally fast, since direct fitting is replaced by fitting  two linear coefficients in the spectral model

\be
  \epsilon(E,T)\approx a_1 \epsilon(E,T_1)+a_2\epsilon(E,T_2),
  \label{eq:2comp}
\ee
where $T_1$ and $T_2$ are the (preset) temperature values which bracket the expected range of
temperature variations in a given cluster and $a_1$ and $a_2$ are the two free
parameters of the model. Reference models $\epsilon(E,T_1)$ and $\epsilon(E,T_2)$ are generated using the Astrophysical Plasma Emission Code (APEC) \citep{APEC}, fixing the Galactic  hydrogen column density to $N_{H} = 9 \times 10^{19}$  cm$^{-2}$ \citep{Dickey.Lockman.1990,2005A&A...440..775K}. 
Once the maps of $a_1$ and $a_2$ are generated, they can be
(adaptively) smoothed and combined to determine the value of the temperature, which
in the simplest form is
\begin{eqnarray}
T=\frac{a_1T_1+a_2T_2}{a_1+a_2}.
\end{eqnarray}
Details and a more accurate expression for the temperature are given in  \cite{Churazov.et.al.1996}. Despite the simplicity of the method, the derived values of temperature show good agreement with the results of direct spectral fitting of individual regions.
We present the (projected) temperature map of the Coma cluster derived from \xmm data using this approach
in the bottom row of Fig.~\ref{fig:all}. 
Each value of the temperature was calculated using adaptively chosen regions containing $\sim 3000$ counts. For this number counts, the expected statistical uncertainty on the ICM temperature is $\lesssim 1$~keV (see Fig.~\ref{fig:tmap_uncertainty}), although systematic uncertainties, associated with the non-X-ray background could be substantial.  The bracketing temperature values are $T_1 = 4$ keV and $T_2 = 10 $ keV. Point sources were removed before the analysis.

%We estimated uncertainties on the temperature determination for 
%different plasma temperatures as a function of photon counts (Fig.~\ref{fig:tmap_uncertainty}).
%These uncertainties should be considered as lower limits as they calculated for an idealized case %without background. 

%xmm\_tsmo\_30.job.ok:
%3000 90. 1.05  30. - c\_goal[counts],size\_max["],dev\_max[shift/size],dev\_max\_asec
%kT =4,7,10 keV

\section{The NGC~4839 group}
\label{sec:ngc4839}

\subsection{Observations and analysis}

\begin{figure*}
\centering
    %\begin{subfigure}[]{}
        \centering
        \includegraphics[clip=t, width=0.8\columnwidth]{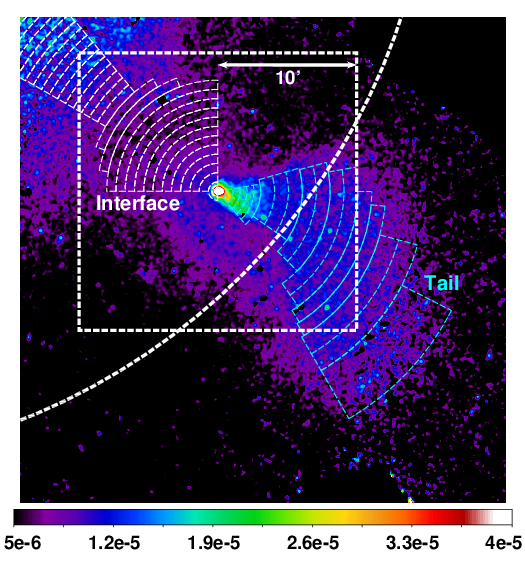}
        \centering
        \includegraphics[clip=t, width=0.8\columnwidth]{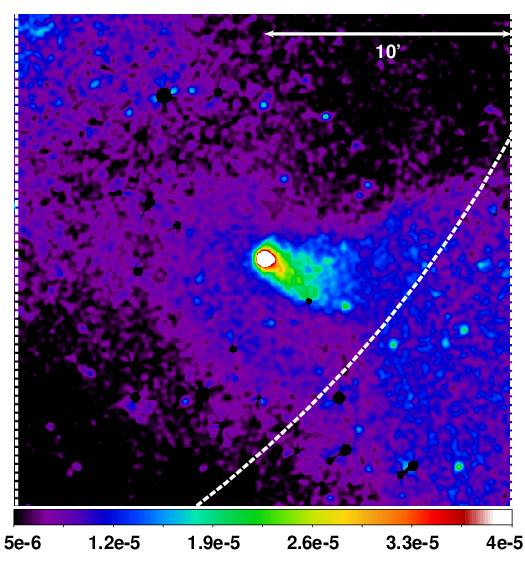}
        \centering
        \includegraphics[clip=t, width=0.8\columnwidth]{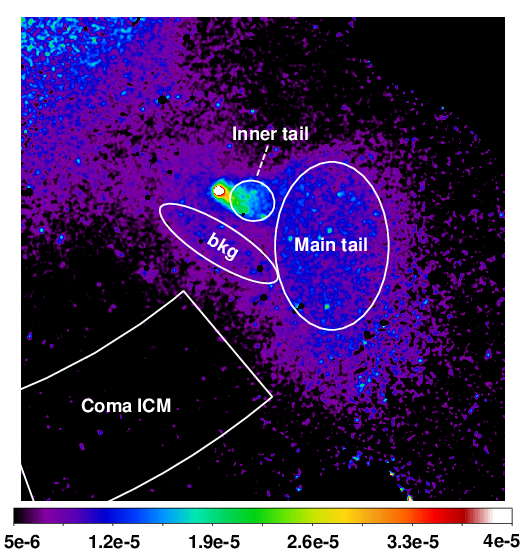}      
        \centering 
        \includegraphics[clip=t, width=0.8\columnwidth]{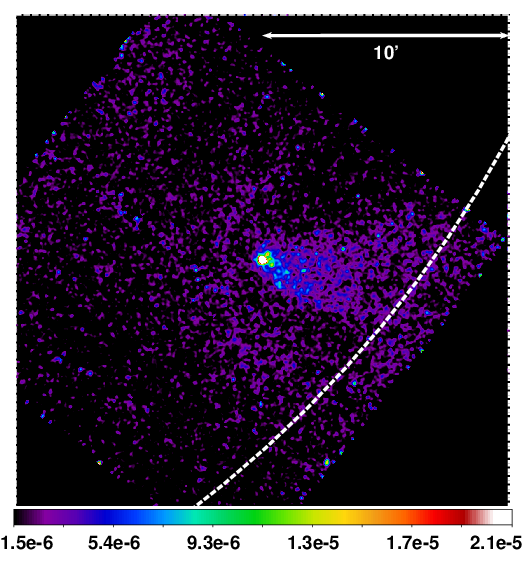}
\caption{\textit{Upper left:} \xmm image (0.5-2.5 keV) of the NGC~4839 group showing the elongated tail and the enhanced X-ray emission toward the direction of the Coma cluster (north-east of the NGC~4839 nucleus). Point sources were excluded. The dashed circle marks $r_{500}$. The annular sectors are used in the analysis discussed in Sections~\ref{sec:tail} and\ref{sec:interface}. 
\textit{Upper right: } The zoom-in view of the rectangular region marked with a white  box ($20'\times 20'$) in the left panel. The white scale bar indicates 10~arcmin.
\textit{Bottom left:} The same image as in the upper panel, but showing regions discussed in Section~\ref{sec:discuss}.
\textit{Bottom right:} \chandra image (0.5-2.5 keV) covering the same region as in the upper right panel.}
\label{fig:chandra_and_xmm}
\end{figure*}

For the analysis of the NGC~4839 group, we use available archival \xmm observations which cover the group and its vicinity.
Some of the pointings - namely, 0652310201, 0652310301, 0652310501, 0652310601, and
0652311001 - were strongly affected by flares and, as a consequence, excluded from our analysis. 
From other observations, flares were removed by discarding all 15-second long time bins which contain more than 6 counts with energy above 10~keV per bin.
The total effective exposure (see Table~\ref{tab:sample}) of all observations (MOS data\footnote{since we are looking at low surface brightness regions, we restrict ourselves to MOS data with a lower and more stable background compared to PN data}) used in the analyses in this Section is 283 ks, and for those centered on the NGC~4839 group, the exposure is 114 ks. 
Fig.~\ref{fig:chandra_and_xmm} (all panels except the lower right) shows the background-subtracted, exposure- and vignetting-corrected \xmm image in the 0.5-2.5 keV energy band. % smoothed by a Gaussian ... .
Point sources were excluded to highlight the diffuse emission SW from the core of the NGC~4839 group (tail) and toward the direction of the Coma cluster center. 
Regions labeled as `Tail' and `Interface' (Fig.~\ref{fig:chandra_and_xmm}, upper left panel) are analyzed in sections~\ref{sec:tail} and~\ref{sec:interface}, regions `inner tail', `main tail', `bkg' and `Coma ICM' are discussed in Section~\ref{sec:discuss}.     

\begin{table*}
\centering
\caption{\label{tab:obsid} Summary of the \xmm observations used for the analysis. Columns list the observation ID, RA-DEC coordinates, offset from NGC~4839 in arcmin, observation date, total exposure time, exposure time after cleaning, and total effective exposure time.}
\begin{tabular}{cccccccccc}
\hline
Obs ID & RA & DEC & Offset, & Observation & Total & Flare-filtered & Total eff.  \\
       &    &     & arcmin  & date        & duration, s &  time, s & exposure\\
\hline
%0652310201    & & 5490 \\ highly flared
%0652310301    & & 6090 \\  highly flared
0652310401   & 12h 57m 24.29s & +27d 29$'$ 52.0$''$ & 0.0138 & 2010-06-24 & 23853 & 14325 &   &\\
0652310701   & 12h 57m 24.29s & +27d 29$'$ 52.0$''$ & 0.0138 & 2010-06-16 & 21839 & 10050 &    &\\
0652310801   & 12h 57m 24.29s & +27d 29$'$ 52.0$''$ & 0.0138 & 2010-12-03 & 16915 & 9187.5 &    &\\
0652310901   & 12h 57m 24.29s & +27d 29$'$ 52.0$''$ & 0.0138 & 2010-12-05 & 16919 & 11257.5 &  114.2 ks  &\\
0691610201   & 12h 57m 24.65s & +27d 29$'$ 42.7$''$ & 0.1710 & 2012-06-02 & 37919 & 37327.5 &   &\\
0691610301   & 12h 57m 24.65s & +27d 29$'$ 42.7$''$ & 0.1710 & 2012-06-04 & 35916 & 32092.5 &    &\\
\hline
0124710101   & 12h 56m 47.68s & +27d 24$'$ 07.0$''$ & 9.9641 & 2000-06-21 & 41505 & 34687.5 &   & \\
0124710301   & 12h 58m 32.19s & +27d 24$'$ 12.0$''$ & 16.0817 & 2000-06-27 & 28616 & 18375 &   & \\
0124712201   & 12h 57m 42.51s & +27d 43$'$ 38.0$''$ & 14.3396 & 2000-12-09 & 27592 & 26737.5 & 168.6 ks&\\
0403150101   & 12h 57m 42.51s & +27d 19$'$ 09.7$''$ & 11.4405 & 2006-06-14 & 54415 & 43725 &   & \\
0403150201   & 12h 57m 42.51s & +27d 19$'$ 09.7$''$ & 11.4405 & 2006-06-11 & 55212 & 45052.5 &   & \\
\hline
total             &                &                &         &            &  360.7 ks  & 282.8 ks       &   & \\
\end{tabular}
\label{tab:sample}
\end{table*}

The NGC~4839 group was also observed with the ACIS-I detector on-board the \chandra X-ray observatory in very faint (VFAINT) mode (ObsID 12887), with a total exposure time of 43 ks.
The initial data processing is done using the recent calibration data and following the  procedure described in \cite{Vikhlinin.et.al.2005}. This includes filtering of high background periods, application
of the calibration corrections to the detected X-ray photons, and determination of the background intensity
in each observation. The right panel of Fig.~\ref{fig:chandra_and_xmm} shows the available \chandra observation of the NGC~4839 group. Since the \chandra observation is relatively short and covers only the core of the group and the inner tail, only \xmm data were used for the  spectral analysis.

\subsubsection{The tail of NGC~4839}
\label{sec:tail}

\begin{figure}
	\includegraphics[width=\columnwidth]{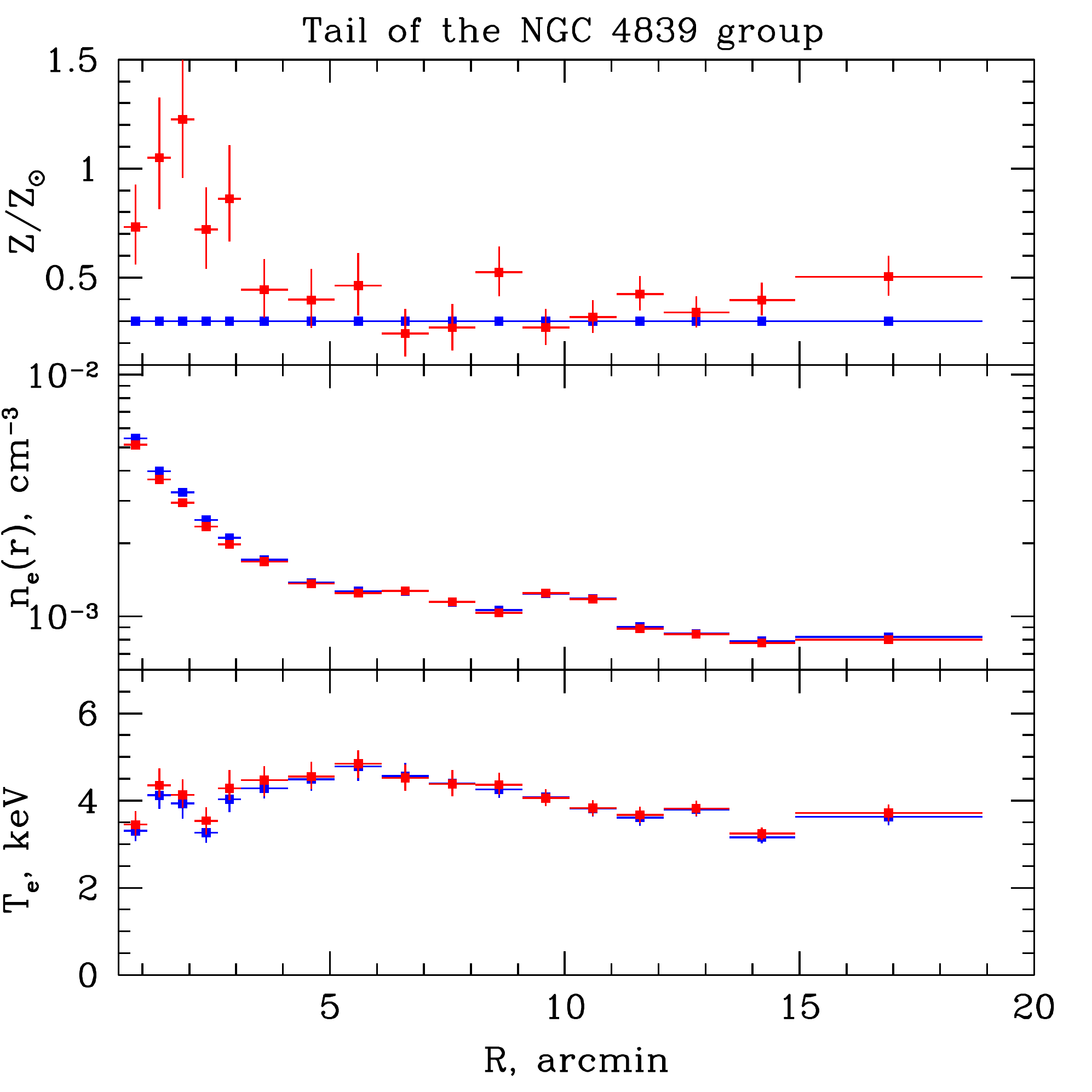}
    \caption{Radial temperature and electron density profiles along the tail of the NGC~4839 group (see regions in Fig.~\ref{fig:chandra_and_xmm}, upper left panel). In blue, we show the results of spectral fitting with the ICM abundance fixed at 0.3 solar, while the case with varying gas metallicity is shown in red.}
    \label{fig:tail}
\end{figure}

For spectral analysis of the tail of the NGC~4839 group, we again employ the model PHABS $\times$ APEC, i.e. 
an absorbed single temperature thermal plasma model. We adopted the same value of the Galactic hydrogen column $N_{H} = 9 \times 10^{19}$  cm$^{-2}$ as in Section~\ref{sec:temp}.  
The spectral fitting is performed with \textit{XSPEC} 12.9.1 for a set of regions labeled as `Tail' in the top-left panel of Fig.~\ref{fig:chandra_and_xmm}.  Fig.~\ref{fig:tail} (lower panel) shows the resulting projected temperature profiles with the gas metallicity\footnote{relative to the solar values of \cite{Lodders}} fixed at 0.3 solar \citep{Simionescu.et.al.2013} and for a varying abundance.  The tail of the group extends behind NGC~4839 out to $\simeq 600$ kpc  and its (projected) temperature is essentially flat at $\simeq 4$ keV (consistent within the uncertainties  with the results of \citealt{Akamatsu.et.al.2013} and \citealt{Sasaki.et.al.2016}). The middle panel of Fig.~\ref{fig:tail} shows the electron density profile obtained from the normalization $K$ of the APEC model:

\be
K = \frac{10^{-14} n_e n_H \delta V}{4 \pi (D_A(1+z))^2},
\label{eq:norm}
\ee
where $D_A$ is the angular diameter distance to the Coma cluster, $z = 0.0231$ is  Coma's redshift, $n_e$ and $n_H \simeq n_e/1.2$ (for fully ionized plasma) are the mean electron and hydrogen number densities, respectively, and $\delta V$ is the volume of the spectrum extraction region. We approximate the tail as a set of conic sections and assume that the size of each section along the line of sight is equal to its transverse dimension. 
The gas mass density is $\rho_{gas} = \mu m_p (n_e + n_H)$ with $\mu = 0.61$ being the mean molecular weight; and the gas mass of the tail $M_{gas, tail} = \int \rho_{gas} dV \simeq 1 \times 10^{12} M_{\odot}$. While statistical uncertainties in the gas mass are small ($\simeq 0.1 \times 10^{12} M_{\odot}$), the uncertainty associated with the geometry and size of the tail could change the gas mass estimate by a factor of $\sim 2$.      
The derived $M_{gas, tail}$ agrees well with that based  on {\it Suzaku} data \citep{Sasaki.et.al.2016}: $M_{gas, tail} \simeq (0.96 \pm 0.03) \times 10^{12} M_{\odot}$.

\subsubsection{Interface between Coma and the NGC~4839 group}
\label{sec:interface}

\begin{figure}
	\includegraphics[width=\columnwidth]{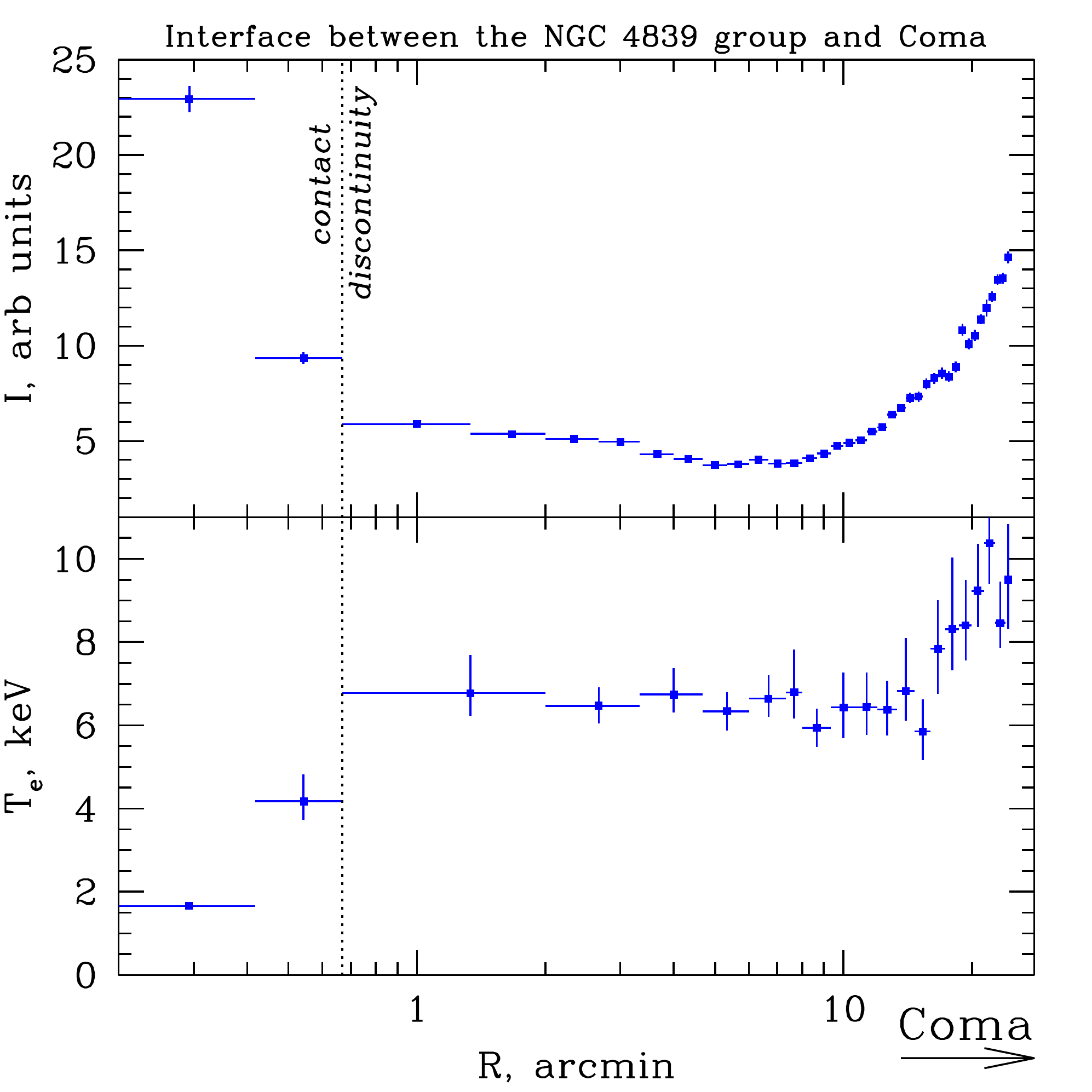}
    \caption{ `Interface' region between the NGC~4839 group and the Coma cluster. The profiles are centered on NGC~4839 (see Fig.~\ref{fig:chandra_and_xmm}, upper left) with small radii ($\lesssim 1'$) lying in the core of the group and larger radii ($\gtrsim 7-8'$) corresponding to the main Coma cluster. Surface brightness and temperature profiles result from the spectral fitting of an absorbed  single-temperature thermal plasma model (phabs x apec). The exact regions are labeled as `interface' in Fig.~\ref{fig:chandra_and_xmm} (upper left panel).} 
    \label{fig:bridge}
\end{figure}

As mentioned in the introduction, \cite{Colless.Dunn.1996} argued that the merger axis lies almost in the plane of the sky ($\alpha \simeq 75^{\circ}$) with a  velocity of 1700 km s$^{-1}$ which corresponds to a Mach number $M = 1.5$ for the ambient gas temperature $\sim$5 keV.
In this scenario, one expects a bow shock to form ahead of the infalling group.    
From the Rankine-Hugoniot jump conditions \citep{LL1987}, the expected density and temperature jumps can be expressed as a function of the Mach number: 
\begin{equation}
\frac{{\rho}_2}{{\rho}_1}=\frac{\mathcal{M}^2(\gamma+1)}{2+\mathcal{M}^2(\gamma-1)},
\end{equation}
\begin{equation}
\frac{T_2}{T_1}=\frac{[(\gamma-1)\mathcal{M}^2+2][2\gamma \mathcal{M}^2-(\gamma-1)]}{(\gamma+1)^2\mathcal{M}^2}.
\end{equation}
For a Mach number $M \simeq 1.5$ and $\gamma = 5/3$,  the expected density jump 
is $\rho_2/\rho_1\simeq 1.7$ and the 
expected temperature jump is $T_2/T_1 \simeq 1.5$. From the X-ray images (see Fig.~\ref{fig:chandra_and_xmm}), it is tempting to associate a `sheath' region  (the region of enhanced X-ray surface brightness between the NGC~4839 core and Coma) with the shocked gas. The edge of the `sheath' region is located  $R \sim 3'-4'$  away from the group center. At these radii the projected temperature is  $T \sim 6$ keV.  
To test this hypothesis, we extracted spectra for the set of regions shown in Fig.~\ref{fig:chandra_and_xmm} (top left) and labeled as `Interface'. 
Fig.~\ref{fig:bridge} shows the surface brightness (APEC normalization) and projected temperature profiles. 
A steep decline of the surface brightness and a steep temperature increase at small distances ($ R \lesssim 1'$) from the NGC~4839 core clearly correspond to a cold front (a contact discontinuity, separating cold gas in the core from the hotter ambient ICM).  Beyond this contact discontinuity, the surface brightness demonstrates a gradual decline (at $ R \simeq 1.5' - 8'$) and a subsequent increase (at $R \gtrsim 8'$). While the latter is clearly associated with the Coma ICM, the nature of the former is unclear, since we do not see clear signatures of a shock in the data shown in   Fig.~\ref{fig:bridge}. 
We return to this question in the next sections, where we run a series of numerical simulations to  explain the observed profiles and to place constraints on the merger scenarios.

\section{Simulation methods}
\label{sec:setup}

We performed smoothed-particle hydrodynamic (SPH) simulations to investigate the merging scenario of NGC 4839 group by using the Gadget-2 code \citep{Springel2001}. To simplify the merger process, we only consider the interaction between a main cluster and an infalling subcluster. Each body is further modelled as a spherical object consisting of a dark matter (DM) halo and gas atmosphere. A detailed description of the simulation method has been given in \citet{Zhang2014,Zhang2015}. Here, we only provide a brief summary.
\begin{itemize}
   \item The merging process is modelled in Cartesian coordinates $(x,\ y,\ z)$. The center of mass of the merging system is initially set at rest at the origin of the coordinates. The merger plane coincides with the $x-y$ plane.  The DM halos and gas halos follow the Navarro-Frenk-White (NFW) profile \citep{Navarro1997} and the Burkert profile \citep{Burkert1995} within the virial radius for the density profiles (see equations~(1)--(4) in \citealt{Zhang2014}). We fix the masses of halos as described below and set  the concentration parameter of the main cluster $c=4$ according to the weak-lensing measurement \citep{2014PASJ...66...99O}. The concentration parameter of the subcluster, however, is determined from the mass-concentration relation of \cite{Duffy2008}.
   \item The merger configuration between two clusters is described by the following four parameters: the virial mass  of the main cluster $M_{200}$, the mass ratio between the main cluster and the subcluster $\xi\ (>1)$, the initial relative velocity $V_0$, and the impact parameter $P_0$. Motivated by recent weak lensing studies \citep{2014PASJ...66...99O}, we fix the Coma cluster mass at $M_0=1.2\times10^{15}\msun$ in all our simulations, but vary the other three parameters to find a `best-fit' model for NGC 4839 (see Table~\ref{tab:simulation_parameters} for the parameter settings used in the simulations).
    We consider two different values for the mass ratio between Coma and the NGC~4839 group: $\xi = 10$ and $60$. The former corresponds to the mass ratio estimate of \cite{Colless.Dunn.1996}, and the latter is motivated by the gas mass estimate obtained in Section~\ref{sec:tail}:
if we assume the baryon fraction of $\sim 10\%$, then  the total mass of the NGC~4839 group is $\sim 10^{13} M_{\odot}$. We return to the question of the mass ratios below.
     
\end{itemize}

We stress here that, we do not intend to find an exact match between our simulations and observations in this study, but rather  to understand the possible merger scenario(s). That is the reason why we did not survey a large parameter space, but test a few specified merger cases with large/small mass ratios and high/low initial angular momentum. We also only present the simulation results while assuming the line of sight (LOS) is parallel to the $z$-axis, since, as mentioned above, available estimates of the viewing angle \citep{Colless.Dunn.1996} favour a plane of the sky merger. Moreover, if the group is on its first radial infall into the Coma cluster, as it is widely assumed, the group is expected to move supersonically with an infall velocity of $\sim 2000 $ km s$^{-1}$.    
The line-of-sight velocity of the NGC~4839 group relative to the Coma cluster is only $V_r \simeq 470$ km s$^{-1}$ \citep{Adami.et.al.2005}, i.e. much smaller compared to the local speed of sound. Thus, to reconcile the measured LOS velocity with the first radial infall, the viewing angle should be close to $90^{\circ}$.

\begin{table*}
\begin{center}
\caption{Merger parameters of SPH simulations.}
\label{tab:simulation_parameters}
\begin{tabular}{ccccc}
\hline \hline
ID & $M_{200}\ (10^{15}\msun)$ & $\xi$ & $V_0\ (\kms)$ & $P_0\ (\kpc)$ \\ \hline
R10V500P2000 & $1.2$ & 10 & $500$ & $2000$ \\ \hline
R10V1000P4000 & $1.2$ & 10 & $1000$ & $4000$ \\ \hline
R60V500P2000 & $1.2$ & 60 & $500$ & $2000$ \\ \hline
R60V1000P3000 & $1.2$ & 60 & $1000$ & $3000$ \\
\hline\hline
\end{tabular}
\end{center}
\end{table*}

\section{Results}
\label{sec:results}

We discuss two possible merger scenarios for the NGC 4839 group in this section, including (1) the group is in the pre-merger stage (i.e. before the primary pericentric passage\footnote{For convenience, we set the evolution time $t=0$ at the moment of primary pericentric passage.}); (2) the group is near the primary \mbox{apocenter}. We show that our SPH simulations favour the latter scenario, but the former one still cannot be definitely excluded.

\subsection{Pre-merger scenario} \label{sec:results:pre-merger}

\begin{figure}
	\centering
	\includegraphics[width=0.95\columnwidth]{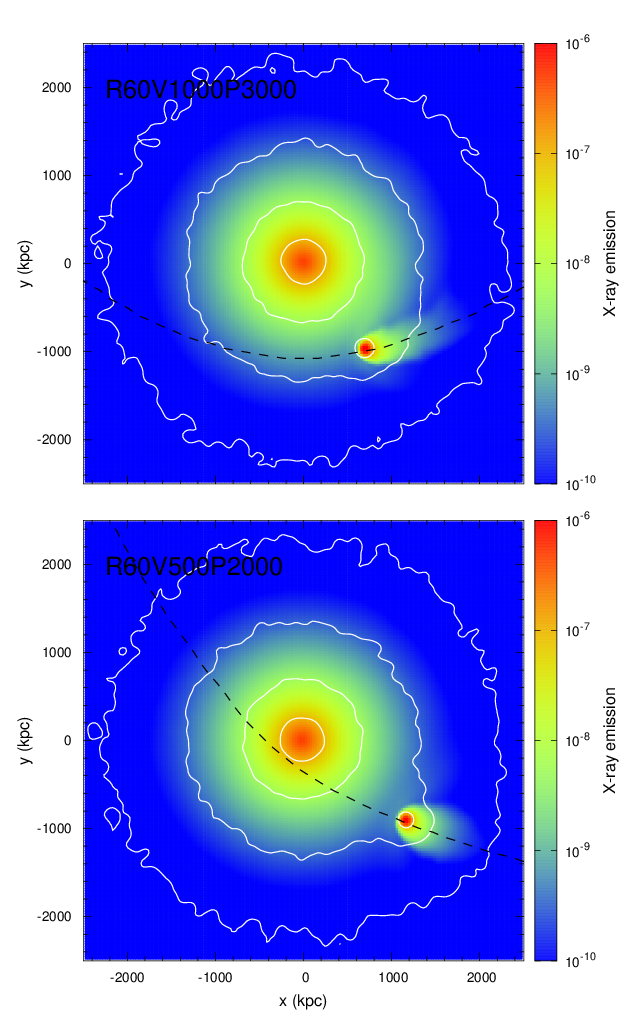}
	\centering
\begin{tikzpicture}
    \draw (0, 0) node[inner sep=0] {\includegraphics[width=0.95\columnwidth]{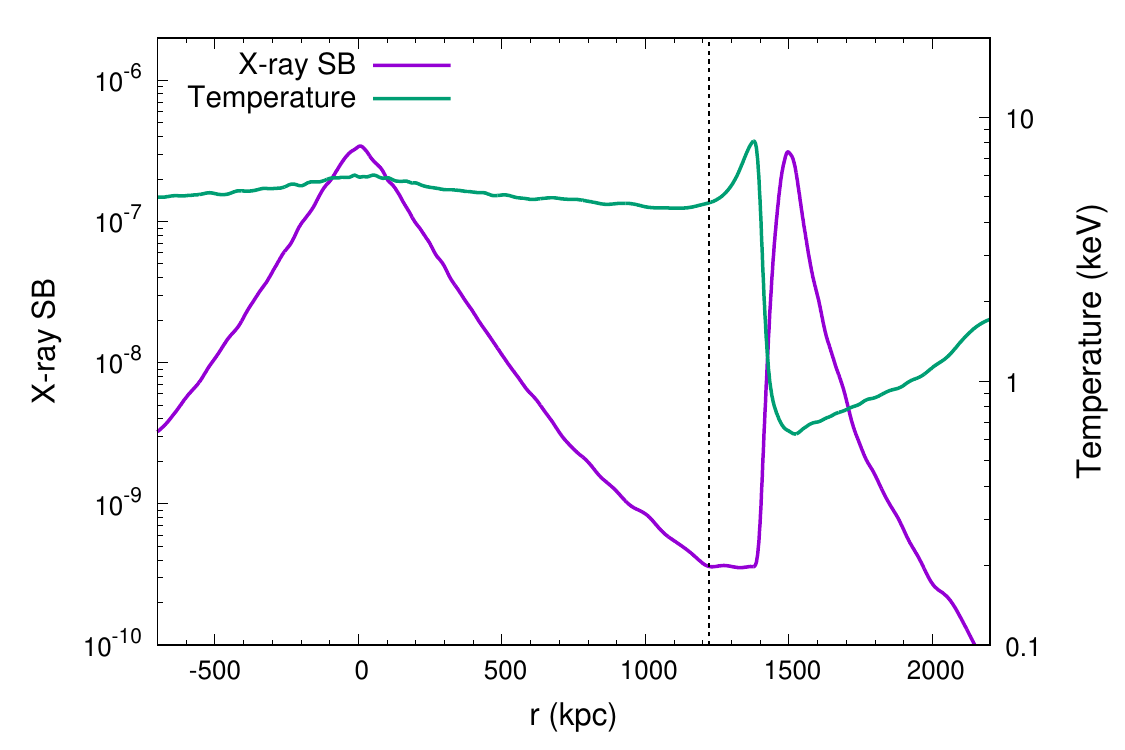}};
	\draw (-1.7,2.7) node {R60V500P2000};	
\end{tikzpicture}	
\caption{\textit{Upper and middle panels:} X-ray surface brightness distribution of the runs  R60V1000P3000 and R60V500P2000 at $t=-0.5\Gyr$. The white contours show the distribution of the mass surface density. The black dashed curves reveal the trajectory of the infalling subcluster.  
\textit{Lower panel:} profiles of the X-ray surface brightness and X-ray emission weighted temperature (for R60V500P2000) along the line connecting the centers of the two merging subclusters. The horizontal axis represents the distance from the center of the main cluster. Note that the tail is symmetric and the bow shock is not prominent in the X-ray surface brightness map/profile and is actually located in the region of minimal surface brightness (for details see the text).  }
\label{fig:pre_merger_map}
\end{figure}

Fig.~\ref{fig:pre_merger_map} (top and middle panels) shows the X-ray surface brightness distribution in the runs R60V500P2000 and R60V1000P3000 at $t=-0.5\Gyr$, where the dashed line shows the trajectory of the infalling subcluster throughout the simulation. In this image, we  see a straight gaseous tail trailing  the subcluster. The subcluster, however, still retains most of the gas in its gravitational potential well, because it has not yet entered the high gas density region in the main cluster. 
Therefore, in this scenario, the gas mass of the tail $\sim 10^{12} M_{\odot}$ (see Section~\ref{sec:tail}) can be used to estimate the total mass of the group $M_{200} \sim 10^{13} M_{\odot}$, assuming a gas mass fraction of $\sim 10\%$ \citep[][]{2006ApJ...640..691V,2009ApJ...693.1142S}.
 Fig.~\ref{fig:pre_merger_map} (lower panel) shows the profiles of the X-ray surface brightness and the X-ray emission weighted temperature along the line connecting the centers of the two clusters. The vertical dashed line in this panel marks the position of the shock front. Notice that the shock front itself  is not prominent in the X-ray surface brightness profile. 
Typically, shock fronts are detected in X-ray images as surface brightness edges. Our simulations suggest     
that the shock front, associated with the NGC~4839 group, is actually located close to the lowest surface brightness region in the X-ray image (Fig.~\ref{fig:pre_merger_map}) and not in a region of surface brightness enhancement. 
Motivated by these results, one can expect to detect a bow shock at a distance of $\simeq 5-6 $ arcmin (the lowest surface brightness region in Fig.\ref{fig:bridge}) from the contact discontinuity (cold front), corresponding to the leading edge of the group.   
At the same time, the stand-off distance $\Delta$, i.e. the distance between a stagnation point and the closest point on the bow shock, can be estimated from the Mach number \citep[][among others]{Moekel1949, Farris1994, Verigin2003, 2018arXiv180802885Z}. 
According to  equation (35) in \cite{Verigin2003} (see also equation (A4) in \citealt{2018arXiv180802885Z}), for the  NGC~4839 group, infalling radially and with a Mach number $\simeq 1.5$, the stand-off distance $\Delta \simeq 0.7 \times R_{cf}$, where $R_{cf} \simeq 1'$  is the curvature radius of the cold front. Thus, we have a clear contradiction between predictions for the bow shock position. Projection effects do not play a significant role here, since, as discussed above, the merger should happen almost in the plane of the sky if the NGC~4839 group is on its first infall into Coma.  So we conclude that available X-ray data and our calculations of the stand-off distance disfavour the pre-merger stage.

Moreover, since the stripped gas is distributed mostly along the trajectory of the infalling subcluster before the core passage,  runs R60V500P2000 and R60V1000P3000 show different tail directions relative to the center of the main cluster if the LOS is perpendicular to the merger plane (see Fig.~\ref{fig:pre_merger_map}). These two runs have different initial angular momenta. In practice, however, there is a degeneracy while determining the angular momentum of the subcluster and the viewing angle from the \textit{Chandra}/\textit{XMM-Newton} X-ray images (the wake of NGC~4839 is nearly parallel to the radial direction of Coma in the sky plane). In spite of this, tails of the subclusters remain almost straight and symmetric in both runs, because there is no sharp turn in their motion (it only occurs near the \mbox{apocenter} when the merging system has a relatively small initial angular momentum, see Section~\ref{sec:results:post-merger} for more discussion). This is obviously different from what we observe in NGC~4839. In this regard, our simulations disfavour the pre-merger scenario. However, we stress that it is still possible to explain the observed `wiggling' tail as a result of von K\'{a}rm\'{a}n vortex shedding, which generally occurs behind a moving blunt body with a Reynolds number $\gtrsim100$ \citep[e.g.,][]{Williamson1996}. The onset of vortex shedding, however, is usually delayed in numerical simulations since numerical viscosity suppresses  growth of the instabilities \citep{Braza1986}. We simply estimate the vortex shedding period $T_{\rm vortex}$ of the wake if assuming the Strouhal number ${\rm St}=D/T_{\rm vortex}U\simeq0.2$, where $U\ (\sim2000\kms)$ and $D\ (\sim200\kpc)$ are the velocity and size of the infalling cluster in the simulation. The period $T_{\rm vortex}\sim0.5\Gyr$ is thus shorter than (or comparable to) the crossing time (size of the subcluster divided by the velocity of an oncoming stream of the surrounding gas)  of the subcluster.

\subsection{Post-merger scenario} 
\label{sec:results:post-merger}

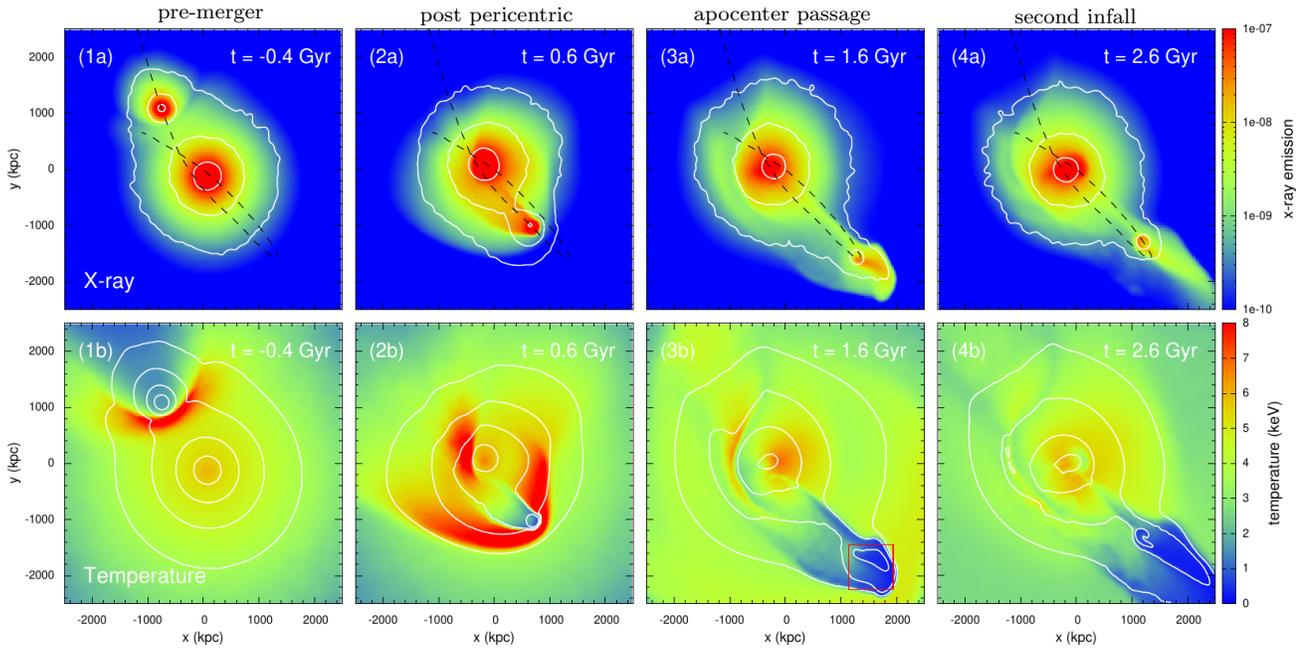
\begin{figure*}
\begin{tikzpicture}
    \draw (0, 0) node[inner sep=0] {\includegraphics[width=0.99\textwidth]{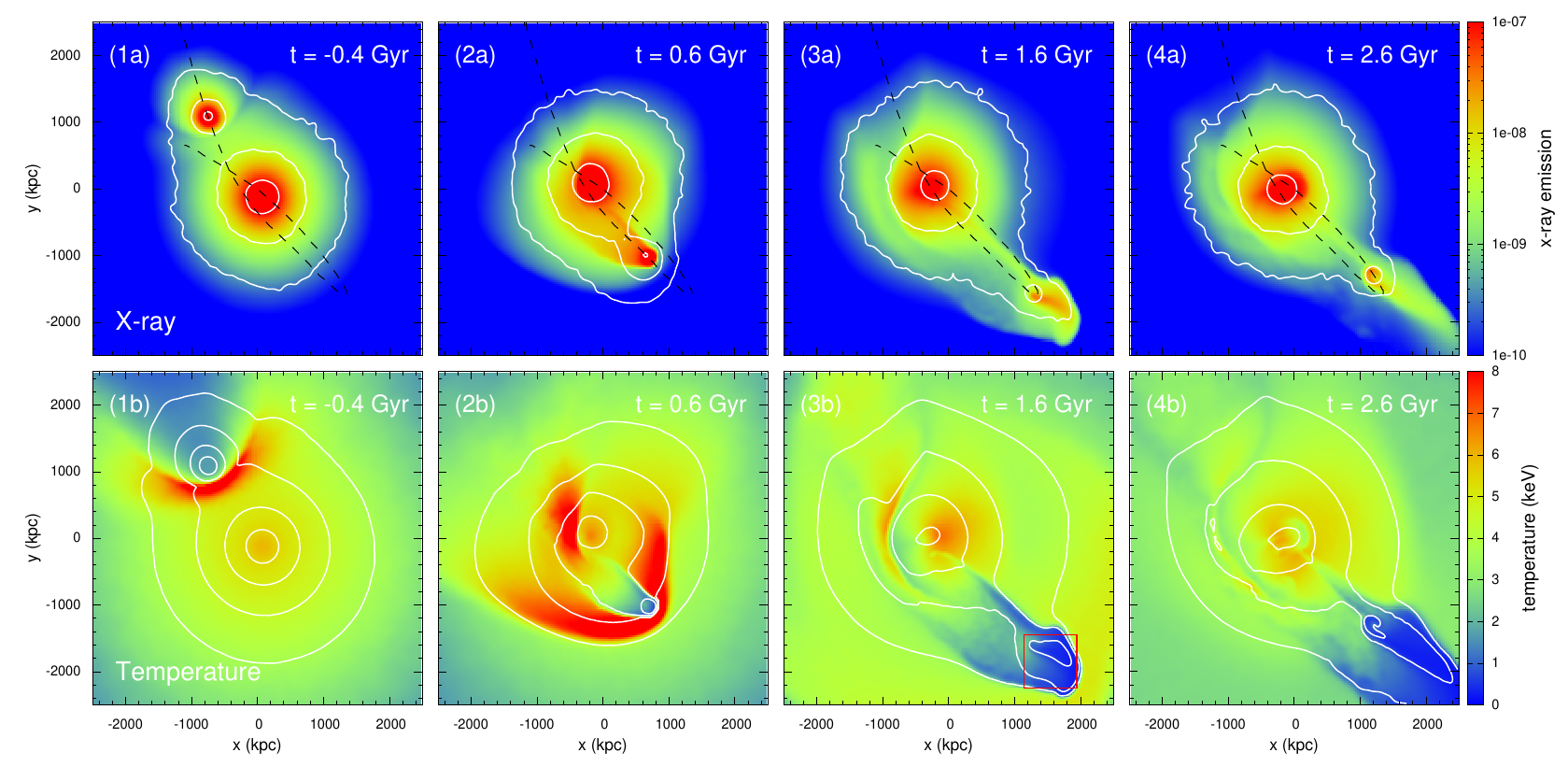}};
	\draw (-5.8,4.3) node {pre-merger};
	\draw (-2,4.3) node {post pericentric};
	\draw (1.8,4.3) node {\mbox{apocenter} passage};
	\draw (5.7,4.3) node {second infall};
\end{tikzpicture}
\caption{Time evolution of the X-ray surface brightness (top row) and X-ray weighted temperature (bottom row) distributions of the run R10V500P2000. These panels illustrate four stages of the merger process (from left to right): pre-merger, post pericentric passage, apocentric passage, and secondary core accretion. The white contours show the mass surface density (top panels) and the X-ray surface brightness (bottom panels) distributions. The black dashed curve in the top panels shows the trajectory of the infalling subcluster with the initial infall located at the top left of the box. The gas mass belonging to the subcluster within the red box marked in panel~(3b) is $M_{\rm gas}\simeq5\times10^{12}\msun$, broadly consistent with the gas mass estimate from the X-ray analysis. Panel~(3a) provides a good match with the X-ray morphology of the NGC 4839 group, when the subcluster crosses the \mbox{apocenter} (compare with Fig.~\ref{fig:all}), which favours the post-merger scenario. }
\label{fig:xray_evolution}
\end{figure*}

\begin{figure*}
\begin{tikzpicture}
    \draw (-4.6, 0) node[inner sep=0]{\includegraphics[width=0.5\textwidth,clip=t,angle=0.,scale=0.98]{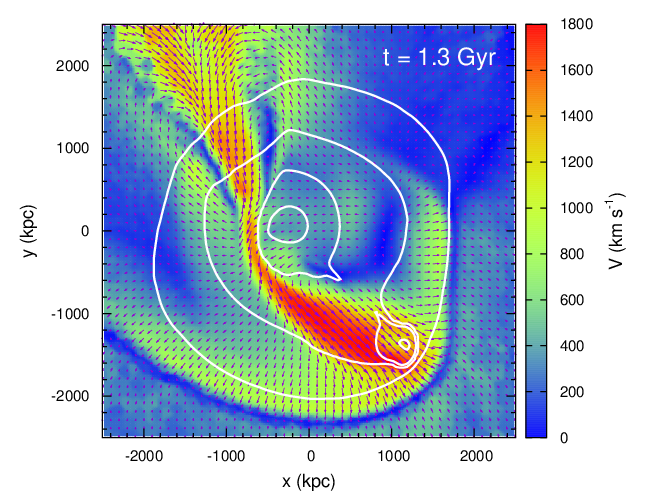}};
    \draw (4, 0) node[inner sep=0]{\includegraphics[width=0.5\textwidth,clip=t,angle=0.,scale=0.98]{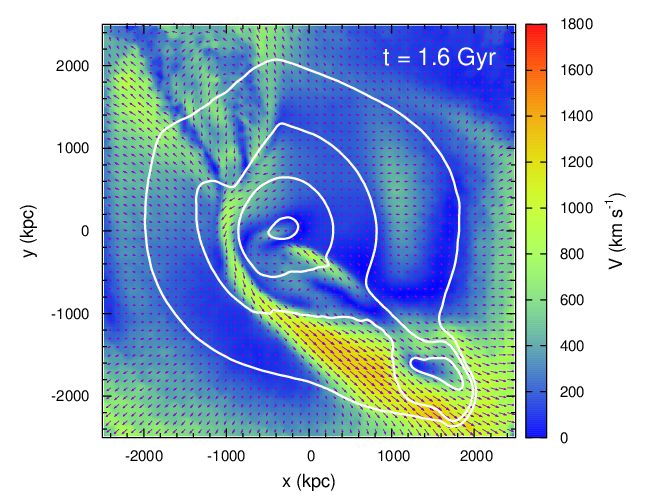}};
	\draw (-4.6,3.5) node {before reaching the \mbox{apocenter}};
	\draw (3.8,3.5) node {apocentric passage};
\end{tikzpicture}
\caption{The slice in gas velocities (taken in the merger plane)  for the post-merger scenario, overlapped with the X-ray surface brightness (white contours). The colormap color-codes absolute values of velocities in the rest frame of the mass center of the merging clusters (Coma + NGC~4839). Arrows show the velocity vectors. The \textit{left} and \textit{right} panels show the moments before and after the \mbox{apocentric} passage, respectively. As the NGC~4839 group approaches the \mbox{apocenter} (left panel), it slows. The ram pressure decreases and the tail gas falls back towards the core of the group, driven by the gravitational drag from the subcluster. Just after the \mbox{apocentric} passage (right panel), the NGC~4839 group starts its second infall into the Coma cluster core, while the gaseous tail is  moving in the opposite direction (toward the south-west). The right panel shows the same time as  panels (3a) and (3b) in Fig.~\ref{fig:xray_evolution}.}
\label{fig:vel}
\end{figure*}

\begin{figure}
	\centering
\begin{tikzpicture}
    \draw (0, 0) node[inner sep=0] {\includegraphics[width=0.95\columnwidth]{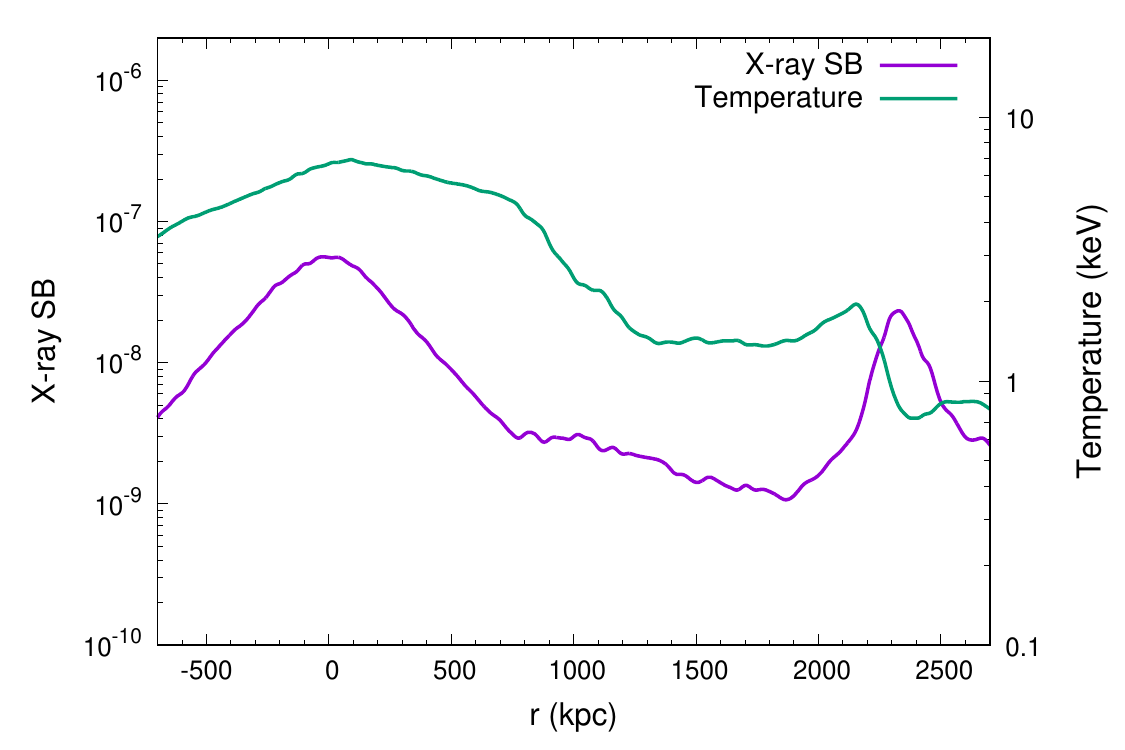}};
	\draw (-1.7,2.7) node {apocentric passage};	
\end{tikzpicture}	
\caption{Same as in the bottom panel of Fig.~\ref{fig:pre_merger_map}, but for the post-merger scenario (R10V500P2000, t = 1.6 Gyr).
The surface brightness and temperature profiles bear some morphological similarities with the pre-merger profiles, but 
the shock between the group and the main cluster is now absent. As a result, no very sharp features are seen in the profiles and the amplitude of temperature variations is markedly smaller.  }
\label{fig:post_merger_map}
\end{figure}

Fig.~\ref{fig:xray_evolution} shows the time evolution of the X-ray surface brightness and X-ray weighted temperature distributions in the run R10V500P2000. 
The panels illustrate four different merger stages, i.e. 1) pre-merger, 2) post pericentric passage, 3) apocentric passage, and 4) secondary core accretion (from left to right). The simulated image shown in Fig.~\ref{fig:xray_evolution} panel~(3a) (apocentric passage)   resembles the observed X-ray morphology (see Fig.~\ref{fig:all} and \ref{fig:chandra_and_xmm}).  Our simulation provides a good match with the real X-ray observations (compare panel~(3a) in Fig.~\ref{fig:xray_evolution} with the upper panel of Fig.~\ref{fig:all}), when the subcluster is close to, but shortly after, the primary \mbox{apocentric} passage. The subcluster trails a large and asymmetric tail, whose formation could be understood in the following way: 
\begin{itemize}
	\item During the pericentric passage, the gas in the NGC~4839 group is driven by the ram pressure and forms a tail trailing the galaxy.
	\item  When the subcluster approaches  \mbox{apocenter}, it slows  and reverses its radial velocity. 
	\item At the same time, the ram pressure decreases and the subcluster gas falls into the local gravitational potential well of the group.
	\item Shortly after the \mbox{apocenter} passage, the dark matter, stars and the gaseous core are moving towards the Coma cluster center, while the displaced gas moves in the opposite direction, forming the structure as in the panel~(3a) of Fig.~\ref{fig:xray_evolution}.  
\end{itemize}
The last two stages are illustrated in Fig.~\ref{fig:vel} showing the flow patterns in and around the infalling group. As argued above, this structure is formed when the subcluster is turning sharply. This is the reason why the run R10V500P2000 gives a better match with the observations than R10V1000P4000 (the infalling subcluster holds larger angular momentum in this run). For the large impact parameter (R10V1000P4000),  there is no strong interaction between the infalling group and the main cluster. After pericentric passage, the core of the subcluster remains roughly round, while X-ray data clearly show an edge-like structure at the head of the group.

Fig.~\ref{fig:xray_evolution} shows that the subcluster penetrates the central gas core of the main cluster and a large fraction of gas is stripped away from its potential well through this process. We measure the gas mass belonging to the subcluster within the red box marked in panel~(3b) (integrated over the LOS), i.e. $M_{\rm gas}\simeq5\times10^{12}\msun$. This result is of the same order of magnitude as the gas mass estimate obtained in Section~\ref{sec:tail}. We note that the post-merger scenario requires a smaller mass ratio (i.e. higher mass of the NGC~4839 group) than that of the pre-merger scenario. Otherwise, after the primary core passage, the subcluster loses almost all its gas.  In panels~(1b) and (2b), prominent shock waves are driven by the infalling subcluster. However, shocks usually move faster than the subcluster after  pericentric passage \citep{2018arXiv180802885Z}.
Thus, we do not expect to see shocks near NGC~4839.  The absence of a shock is also illustrated in Fig.~\ref{fig:post_merger_map} which shows the X-ray surface brightness and X-ray emission weighted temperature for the post-merger scenario, i.e., shortly after the \mbox{apocenter} passage,  along the line connecting the group and the main cluster. No very sharp features are seen in the profiles and the amplitude of temperature variations is markedly smaller compared to the pre-merger case shown in Fig.~\ref{fig:pre_merger_map}, bottom panel.

At the moment when the  subcluster is near the \mbox{apocenter}, the  bow shock associated with the first infall of  NGC~4839  has propagated  south-west much farther away from the core of the group and its current position is roughly consistent with the Coma radio relic \citep{1991A&A...252..528G}. We will discuss the origin of the Coma relic in more detail in Lyskova et al. (in prep.).  The observed   hot `sheath' region, as seen in the X-ray data (see Fig.~\ref{fig:all} and \ref{fig:chandra_and_xmm}), that surrounds the brighter NGC~4839 subcluster core, can be also explained under the post-merger scenario. If the impact parameter of the merger is not too high, then the subcluster, just after the \mbox{apocenter} passage, starts moving  through its own tail of stripped gas (see flow patterns in Fig.~\ref{fig:vel}, right panel). An increased temperature and gas density in the `sheath' region could be due to interaction of the re-infalling subcluster with the stripped gas mixed with the Coma ICM. One possible mechanism here is the `stolen atmosphere' effect (described in \citealt{2018ApJ...865..118S}), when intra-cluster gas, surrounding the subcluster,  is drawn into the NGC~4839 group potential and compressed/heated.  We defer the discussion of the nature of observed `sheath' for future work.

To check results obtained with the SPH simulations, we also ran a FLASH simulation using the same merging parameters as those used in run R10V500P2000. The simulated  surface brightness and temperature maps are consistent with results of the SPH simulations. %We also tried to set initial conditions of the merger using the universal pressure profile for the Coma cluster. Again, results stay unchanged. 
Moreover, \cite{Sheardown2019} inspected a large set of FLASH simulations of idealized binary
cluster mergers and independently reached the same conclusion that the NGC~4839 group is most likely to have passed by the Coma core from the north-east with a small impact parameter and is now on
its next infall.

\section{Discussion}
\label{sec:discuss}
\subsection{Mass of the NGC~4839 group}

\begin{table*}
\centering
\caption{Summary of  mass estimates of the Coma cluster and the NGC~4839 group }
\begin{tabular}{lccccccccc}
\hline
object & study &  $M_{200}, 10^{14} M_{\odot}$ \\
\hline
Coma  & M-T, T=8 keV &   12.6  \\
%Coma & \cite{Planck} & 9.3 \\
Coma & \cite{2014PASJ...66...99O} & 12.0 \\
Coma & \cite{Colless.Dunn.1996} & 12.9 \\
NGC~4839  & M-T, T=4 keV  & 4.2  &  \\
NGC~4839  & M-T, T=2.7 keV  & 2.3  &  \\
NGC~4839  & \cite{2014PASJ...66...99O}  & $> 1$  &  \\
NGC~4839 & \cite{Colless.Dunn.1996} & 0.9 \\
\hline
\end{tabular}
\label{tab:masses}
\end{table*}

As discussed above, the mass of the NGC~4839 group is poorly constrained. 
We briefly summarize here the mass and the mass ratio estimates $\xi = M_{vir}(\rm Coma)/M_{vir}(\rm NGC~4839)$ available in the literature.  
From the analysis of the velocity distributions of the 
cluster/group members,  \cite{Colless.Dunn.1996} 
obtained the virial masses of $\simeq 1.3 \times 10^{15} M_{\odot} $ and $\simeq 8.6 \times 10^{13} M_{\odot}$ for the Coma cluster and the NGC~4839 group, respectively. So their mass ratio is  $\simeq 15$. 
%In Table~\ref{tab:masses}, we sum up all the constraints on the group and the main cluster masses.  

Based on the weak lensing signal,   
\cite{2014PASJ...66...99O} measured the Coma virial mass of $M_{vir}(\rm Coma) = 1.2 \times 10^{15 } M_{\odot}$ and the total mass of the NGC~4839 group within the tidal (truncation) radius    
$M(r < 98$ kpc$) \simeq 1.6 \times 10^{13} M_{\odot} $. This estimate was obtained assuming that the group mass density outside the truncation  radius, $r_t = 98$ kpc, is close to zero. This measurement can  constrain the virial ($M_{200}$) mass of the NGC~4839 group, assuming that the mass profile within $r_t$ has not been modified by the merger. We described 
the mass density with the NFW profile and explored a range of $M_{200}$ and $c_{200}$. The weak-lensing estimate agrees reasonably  well with the group virial mass of $> 1\times 10^{14} M_{\odot}$. The NFW mass profiles with $M_{200} \sim 10^{13} M_{\odot}$ are in more than 3$\sigma$ tension with the weak lensing estimate. In Table~\ref{tab:masses}, we provide a summary of the NGC~4839 group mass (and the mass ratio) estimates. Note that this table is not intended to cover all Coma mass measurements available in the literature.       

The masses also can be estimated via X-ray scaling relations. To convert $M_{500}$ to $M_{200}$, we use the following relation (obtained for the concentration parameter $c = 4$):  $R_{500}/R_{200} = 0.65$ and $M_{500}/M_{200} = 5/2 \times (0.65)^3 = 0.69$. If the NGC~4839 group is characterized by 4~keV gas, then according to the M-T relation \citep{2006ApJ...640..691V,2009ApJ...693.1142S}, its $M_{500} \simeq 2.9 \times 10^{14} M_{\odot}$ and $M_{200} \simeq 4.2 \times 10^{14} M_{\odot}$. For the Coma cluster with its 8~keV gas, the M-T relation gives $M_{200} \simeq 1.26 \times 10^{15} M_{\odot}$. These arguments suggest that the mass ratio between the subcluster and the main cluster is $\simeq 3$. However, such a mass ratio implies a major merger scenario and is disfavoured by our simulations - in this case, the Coma cluster would be dramatically disturbed, and strong shocks, propagating to the north-east and south-west, would be generated. The best-fit post-merger simulation with the mass ratio of $\xi \sim 10$ provides a good match to the  observed morphology of the tail of the group, but the measured ratio of gas temperatures in the tail and in the main cluster do not agree well with simulations. In the simulations, the temperatures of the main cluster and the tail of the subcluster are  $\simeq 5$ keV  and $\simeq 0.5 - 1$ keV, correspondingly, i.e. the temperature ratio is $\sim 5-10$. However, the observed ratio of the X-ray temperatures $\simeq$ 8~keV/4~keV $=2$ is noticeably different.  As a consequence, the mass ratio derived from the M-T relation is several times smaller than in simulations.

\subsection{Temperature of the NGC~4839 group}

\begin{table*}
\centering
\caption{Summary of the spectral fits in different regions shown in Fig.~\ref{fig:chandra_and_xmm}, lower left panel. The uncertainties are 1$\sigma$ confidence level. For the fitting of a thermal plasma, we used the single temperature (phabs x apec) model or two-temperature (phabs x (apec$_1$ + apec$_2$)) model with 464 PHA bins. Since the actual metallicity of gas in the `inner tail' region is not known, for fitting the two-temperature model, we consider two cases:  $Z =0.5Z_{\odot}$ and $0.7Z_{\odot}$. }
\begin{tabular}{lccccccccc}
\hline
region & $K_1, 10^{-5}$ & $kT_1$, keV & $Z_1/Z_{\odot}$ & $K_2, 10^{-5}$ & $kT_2$, keV & $Z_2/Z_{\odot}$ &  $\chi^2$/d.o.f. \\
\hline
`bkg'  & $5.12 \pm 0.08 $ & $6.4\pm 0.2 $ & $0.25^{+0.09}_{-0.08}$  & - & - & - & 372/461 \\[0.15cm]
`inner tail' & $8.34 \pm 0.15 $  & $4.2^{+0.2}_{-0.1} $ & $0.71 \pm 0.07$  & - & - & - & 384/461 \\[0.15cm]
`main tail'  & $6.01 \pm 0.03$       & $\mathbf{ 4.2} \pm 0.1$     & $0.3$ (fixed) & -               & -             & -              &   425/462  \\[0.15cm]
\hline
%inner tail & $8.40 \pm 0.07$  & $4.2\pm 0.1 $ & $0.7$ (fixed)  & - & - & - & 380/476 \\[0.15cm]
`inner tail' & $2.94^{+0.66}_{-0.33}$ & $2.2^{+0.3}_{-0.1}$ & 0.5 (fixed) & $6.06^{+0.32}_{-0.67}$ & $6.4$ (fixed) & $0.25$ (fixed) & 387/461 \\[0.15cm]
`inner tail' & $4.40^{+2.08}_{-1.62}$ & $3.1^{+0.8}_{-0.7}$ & 0.7 (fixed) & $4.27^{+1.72}_{-2.25}$ & $6.4$ (fixed) & $0.25$ (fixed) & 379/461 \\[0.15cm]
`main tail'  & $2.27^{+ 0.33}_{-0.30}$  & $\mathbf{ 2.4} \pm 0.3$   & $0.3$ (fixed)   & $3.84^{+0.30}_{-0.33} $ & $6.4$ (fixed) & $0.25$ (fixed) & 407/461 \\
\hline
\end{tabular}
\label{tab:spectral_details}
\end{table*}

\begin{figure}
\centering
\includegraphics[width=0.99\linewidth]{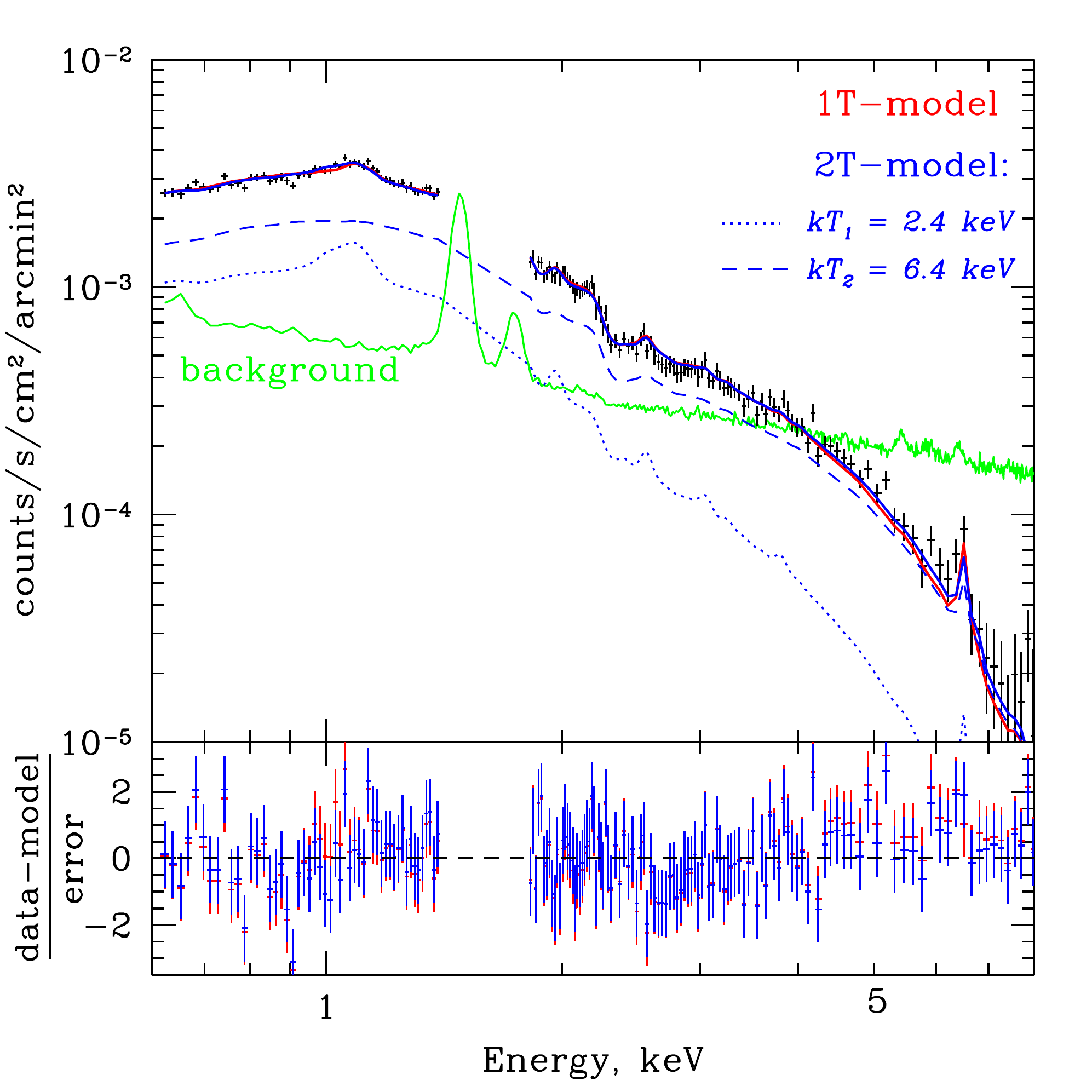}
\caption{The observed spectrum (black crosses) of the `main tail' region and the best-fit models. The single temperature thermal plasma model  is shown in red. The best-fit gas temperature is $4.2 \pm 0.1$ keV (see Table~\ref{tab:spectral_details}). % the residuals from the fit.
A similar quality fit can be obtained with a two-temperature model (blue solid line). In this model, the first component (blue dotted line) is due to the group gas, while the second component (blue dashed line) represents the line-of-sight contribution of the Coma cluster ICM mixed with the stripped group gas.  The temperature and abundance of the second component are fixed at the values obtained for the `bkg' region (see Fig.~\ref{fig:chandra_and_xmm}, lower left panel), viz., $ kT=6.4$ keV and $Z=0.25Z_{\odot}$ (see Table~\ref{tab:spectral_details}). The best-fit temperature of the tail gas decreases  to $2.4 \pm 0.3 $ keV.  The green line shows the background from blank fields to which an additional component (representing a variable particle background in these particular observations) is added. Both the single and two-temperature models fit the spectrum reasonably well (see Table~\ref{tab:spectral_details}). It is therefore plausible that the characteristic temperature of the NGC4389 gas is $\sim 2$ keV, implying a lower initial mass of the group.
}
\label{fig:spectra}
\end{figure}

The projected spectra analyzed in Section~\ref{sec:ngc4839} have not been corrected for the additional contribution that might come from the Coma gas along the line of sight but outside the volume occupied by the tail. This was motivated by the much higher surface brightness of the tail region compared to the typical Coma brightness at the same distance (see the upper panel of Fig.\ref{fig:all}). Even if we re-do our spectral analysis using  the `Coma ICM' region (see Fig.~\ref{fig:chandra_and_xmm}, lower left panel) as a background, the main conclusions remain unchanged. However,  
according to  Fig.~\ref{fig:chandra_and_xmm}, the NGC~4839 group seems to coincide with  some surface brightness enhancement.
For the post-merger scenario, the observed tail is formed when the core of the group turns sharply at the \mbox{apocenter}. So the overall surface brightness enhancement, which was previously thought to be  shocked gas \citep[e.g., ][]{Neumann.et.al.2001}, in our simulations is actually group gas,  stripped before the \mbox{apocenter} passage, and partly mixed with  Coma gas. To account for its contribution to the observed spectra, we treat the region adjacent to the core of the group (marked as `bkg' in Fig.~\ref{fig:chandra_and_xmm}, lower left panel) as the actual background. We  extract and fit the `bkg' spectrum with an absorbed single temperature thermal plasma model. Results are provided in Table~\ref{tab:spectral_details}. 
We model the `main tail' (see Fig.~\ref{fig:chandra_and_xmm}, lower left panel) spectrum with two APEC components. One component is for the `background' gas, the other  is for the tail gas. We fix the temperature and the abundance of the `background' gas at best-fit values from the previous step (see Table~\ref{tab:spectral_details}) and vary its normalization along with  parameters of the APEC model representing the tail gas.
Since the 3D geometry of the merger is poorly known, we allow the background normalization to vary by a factor of $\sim 2$ relative to the  best-fit values for the single temperature fit. As a result, for the two temperature model, the `main tail' temperature decreases to  $\simeq 2.4$ keV. Note that both the single (2 free parameters) and the two-temperature (3 free parameters) models provide an adequate fit to the data (Fig.~\ref{fig:spectra}).  While the modest residuals seen in the bottom panel of Fig.~\ref{fig:spectra} suggest that the spectrum is more complicated than our model, the main point we want to emphasize is that the two temperature model fit with the temperature of one component fixed at 6.4 keV, produces very reasonable normalizations of two components, consistent with an assumption of a cool 2.4 keV group gas embedded into a hotter ICM of the main cluster mixed with the outer layers of the tail. We analyzed also the `inner tail' region (see Fig.~\ref{fig:chandra_and_xmm}, lower left panel) in a similar way. The temperature of the inner tail is found to be $\simeq 2.2$ keV, if we assume that the abundance of heavy elements is $0.5Z_{\odot}$.
A higher metallicity gives higher temperature and larger uncertainties (see Table~\ref{tab:spectral_details}).
If the tail temperature is indeed $2.4$ keV, then the NGC~4839 group mass is $M_{200} \simeq 1.9 \times 10^{14}  M_{\odot} $ from the M-T relation, and the mass ratio between Coma and the group is $\simeq 6.6$. While this doesn't resolve the discrepancy between observations and simulations completely, it certainly reduces the tension.

To some extent the tension between the mass/temperature ratios in simulations and observations also could be weakened  if one finds the best-fit configuration of the merger that exactly matches all available observations, but this task is beyond the scope of our paper.

\section{Conclusions}
\label{sec:conc}

Coma, as one of the nearest massive galaxy clusters, provides a unique close-up view of ongoing
mergers. One of the most striking merger events involves the NGC~4839 group which lies in the Coma cluster outskirts ($\sim 1$ Mpc in projection) in the south-west direction from the cluster center.
The X-ray images of the subcluster exhibit a cold front at the head of the group, a `sheath' region of hotter gas enveloping  the core of the group,  and an elongated tail of ram pressure stripped gas toward
the south-west, i.e. the opposite direction of the Coma cluster center.  
We  discuss two possible scenarios of the merger: (1) the group is on its first infall before the primary pericentric passage, and (2) the group is near the primary \mbox{apocenter}. The data and simulations favour the latter scenario in agreement with the earlier suggestion by \cite{Burns.et.al.1994}.

\begin{itemize}
\item \textbf{Pre-merger scenario}

In the first scenario, the NGC~4839 group comes from the south-west  along the filament connecting Coma with Abell 1367 and it has just started to penetrate the Coma ICM, then the group is expected to move supersonically with  Mach number $ \gtrsim 1.5$. 
The position of the bow shock, predicted by our simulations, corresponds to a surface brightness minimum along the line  
connecting the two merging subclusters, while customarily shock fronts are identified with a sharp increase of the surface brightness.
However, the expected position of a bow shock cannot be reconciled with  the stand-off distance estimate. Thus, we conclude that we  do not see  a strong shock. 
We also showed that if the group is on its first radial infall, then the merger is most likely to be almost in the plane of the sky, i.e. non-detection of the shock cannot be attributed to projection effects. Moreover, the pre-merger simulations do not reproduce the observed X-ray appearance of the tail of the NGC~4839 group. In simulations, tails remain almost straight and symmetric, while the observed morphology is more complex. So we conclude, our analysis disfavours the  radial infall scenario.

\item \textbf{Post-merger scenario}

In the post-merger scenario, a good match between the modelling and the real X-ray observations is achieved when the infalling group has just passed  \mbox{apocenter}. 
Under this scenario, the observed morphology of the tail is formed in the following way. 
When the subcluster approaches the \mbox{apocenter}, it slows  and then reverses its direction of  motion. At the same time, the ram pressure ceases and the group gas falls back into the local potential well of the group, overshooting an equilibrium position and appearing on the other side of the group core. Shortly after  \mbox{apocenter} passage, the core of the subcluster starts moving towards the Coma cluster center, while the displaced gas moves in the opposite direction.
The observed `sheath' - a region of slightly denser and hotter gas than the ambient  gas in front of the group - could arise due to interaction of the subcluster, now moving back towards the Coma center, with its own tail gas now mixed with the gas of the main cluster.  
In the post-merger scenario, we do not expect  shocks near  NGC~4839  since the bow   
shock associated with the first infall of the group  has already propagated towards the south-west, much farther away from the group. The  current position of the shock is roughly consistent with the Coma radio relic. 

\end{itemize}

\section{Acknowledgments} 
The authors thank the anonymous referee for a thorough
review and constructive suggestions which  helped  to  improve  the  paper.
This work was partially supported by the Russian Science Foundation (grant 14-22-00271).
ER acknowledges the support of STFC, through the University of Hull’s Consolidated Grant ST/R000840/1. ER and AS acknowledge access to viper, the University of Hull High Performance Computing Facility. W. Forman and C. Jones acknowledge  support from contracts NAS8-38248, NAS8- 01130, NAS8-03060, the Chandra Science Center, and the Smithsonian Institution. 
The FLASH software used in this work was developed in part by the DOE NNSA ASC- and DOE Office of Science ASCR-supported Flash Center for Computational Science at the University of Chicago.
The scientific results reported in this article are based on observations obtained with \textit{XMM-Newton}, an ESA science mission with instruments and contributions directly funded by ESA Member States and the USA (NASA). This work has also made use of \textit{Chandra} data provided by the Chandra X-ray Center.
This research  has  made  use  of  the  NASA/IPAC  Extragalactic Database  (NED)  which  is  operated  by  the  Jet  Propulsion Laboratory,  California  Institute  of  Technology,  under  contract with the National Aeronautics and Space Administration.

%%%%%%%%%%%%%%%%%%%%%%%%%%%%%%%%%%%%%%%%%%%%%%%%%%%%%%%%%%%%%%%%%%%%%%

\label{lastpage}

\begin{thebibliography}{}

\bibitem[\protect\citeauthoryear{Adami et al.}{2005}]{Adami.et.al.2005} Adami C., Biviano A., Durret F., Mazure A., 2005, A\&A, 443, 17

  
%Suzaku X-Ray Observations of the Accreting NGC 4839 Group of Galaxies and a Radio Relic in the Coma Cluster 
\bibitem[\protect\citeauthoryear{Akamatsu et al.}{2013}]{Akamatsu.et.al.2013} Akamatsu H., Inoue S., Sato T., Matsusita K., Ishisaki Y., Sarazin C.~L., 2013, PASJ, 65, 89 

\bibitem[\protect\citeauthoryear{Andrade-Santos et al.}{2013}]{2013ApJ...766..107A} Andrade-Santos F., Nulsen P.~E.~J., Kraft R.~P., Forman W.~R., Jones C., Churazov E., Vikhlinin A., 2013, ApJ, 766, 107 

  
\bibitem[\protect\citeauthoryear{Arnaud}{1996}]{XSPEC} Arnaud K.~A., 1996, ASPC, 101, 17 

\bibitem[\protect\citeauthoryear{Berrier et al.}{2009}]{2009ApJ...690.1292B} Berrier J.~C., Stewart K.~R., Bullock J.~S., Purcell C.~W., Barton E.~J., Wechsler R.~H., 2009, ApJ, 690, 1292 
  
\bibitem[\protect\citeauthoryear{Biviano et al.}{1996}]{1996A&A...311...95B} Biviano A., Durret F., Gerbal D., Le Fevre O., Lobo C., Mazure A., Slezak E., 1996, A\&A, 311, 95  
  
\bibitem[Braza et al.(1986)]{Braza1986} Braza, M., Chassaing, P., \& Ha, M., H.\ 1986, Journal of Fluid Mechanics, 165, 79
  
\bibitem[\protect\citeauthoryear{Briel et al.}{2001}]{2001A&A...365L..60B} Briel U.~G., et al., 2001, A\&A, 365, L60 

\bibitem[\protect\citeauthoryear{Bonafede et al.}{2009}]{2009A&A...494..429B} Bonafede A., Giovannini G., Feretti L., Govoni F., Murgia M., 2009, A\&A, 494, 429 

%\bibitem[\protect\citeauthoryear{Bonafede et al.}{2009}]{2009A&A...503..707B} Bonafede A., et al., 2009, A\&A, 503, 707 

  
\bibitem[\protect\citeauthoryear{Brown \& Rudnick}{2011}]{Brown.Rudnick.2011} Brown S., Rudnick L., 2011, MNRAS, 412, 2 

\bibitem[Burkert(1995)]{Burkert1995} Burkert, A.\ 1995, ApJL, 447, L25

\bibitem[\protect\citeauthoryear{Burns et al.}{1994}]{Burns.et.al.1994} Burns J.~O., Roettiger K., Ledlow M., Klypin A., 1994, ApJ, 427, L87 
  
  
\bibitem[\protect\citeauthoryear{Cavaliere \& Fusco-Femiano}{1978}]{beta_model} Cavaliere A., Fusco-Femiano R., 1978, A\&A, 70, 677 

  
\bibitem[\protect\citeauthoryear{Churazov et al.}{1996}]{Churazov.et.al.1996} Churazov E., Gilfanov M., Forman W., Jones C., 1996, ApJ, 471, 673

\bibitem[\protect\citeauthoryear{Churazov et al.}{2003}]{2003ApJ...590..225C} Churazov E., Forman W., Jones C., B{\"o}hringer H., 2003, ApJ, 590, 225 


  %%X-ray surface brightness and gas density fluctuations in the Coma cluster
  % beta=0.6, rc=9 arcmin, Rmax simeq 1000 arcsec
\bibitem[\protect\citeauthoryear{Churazov et al.}{2012}]{Churazov.et.al.2012} Churazov E., et al., 2012, MNRAS, 421, 1123 
  

\bibitem[\protect\citeauthoryear{Churazov et al.}{2016}]{Churazov.et.al.2016} Churazov E., Arevalo P., Forman W., Jones C., Schekochihin A., Vikhlinin A., Zhuravleva I., 2016, MNRAS, 463, 1057 

\bibitem[\protect\citeauthoryear{Colless \& Dunn}{1996}]{Colless.Dunn.1996} Colless M., Dunn A.~M., 1996, ApJ, 458, 435 

\bibitem[\protect\citeauthoryear{Di Gennaro et al.}{2018}]{2018ApJ...865...24D} Di Gennaro G., et al., 2018, ApJ, 865, 24 


\bibitem[\protect\citeauthoryear{Dickey \& Lockman}{1990}]{Dickey.Lockman.1990} Dickey J.~M., Lockman F.~J., 1990, ARA\&A, 28, 215 

\bibitem[Duffy et al.(2008)]{Duffy2008} Duffy, A.~R., Schaye, J., Kay, S.~T., \& Dalla Vecchia, C.\ 2008, MNRAS, 390, L64


\bibitem[\protect\citeauthoryear{Dolag et al.}{2009}]{2009MNRAS.399..497D} Dolag K., Borgani S., Murante G., Springel V., 2009, MNRAS, 399, 497
  
\bibitem[\protect\citeauthoryear{Ensslin et al.}{1998}]{1998A&A...332..395E} Ensslin T.~A., Biermann P.~L., Klein U., Kohle S., 1998, A\&A, 332, 395 
  
  
\bibitem[\protect\citeauthoryear{Erler et al.}{2015}]{Erler.et.al.2015} Erler J., Basu K., Trasatti M., Klein U., Bertoldi F., 2015, MNRAS, 447, 2497 


\bibitem[\protect\citeauthoryear{Farris \& Russell}{1994}]{Farris1994} Farris M.~H., Russell C.~T., 1994, JGR, 99, 17 


\bibitem[\protect\citeauthoryear{Geller \& Huchra}{1989}]{1989Sci...246..897G} Geller M.~J., Huchra J.~P., 1989, Sci, 246, 897 


\bibitem[\protect\citeauthoryear{Genel et al.}{2010}]{2010ApJ...719..229G} Genel S., Bouch{\'e} N., Naab T., Sternberg A., Genzel R., 2010, ApJ, 719, 229 


\bibitem[\protect\citeauthoryear{Giovannini, Feretti, \& Stanghellini}{1991}]{1991A&A...252..528G} Giovannini G., Feretti L., Stanghellini C., 1991, A\&A, 252, 528 

\bibitem[\protect\citeauthoryear{Kalberla et al.}{2005}]{2005A&A...440..775K} Kalberla P.~M.~W., Burton W.~B., Hartmann D., Arnal E.~M., Bajaja E., Morras R., P{\"o}ppel W.~G.~L., 2005, A\&A, 440, 775

\bibitem[\protect\citeauthoryear{Kravtsov \& Borgani}{2012}]{2012ARA&A..50..353K} Kravtsov A.~V., Borgani S., 2012, ARA\&A, 50, 353 


\bibitem[\protect\citeauthoryear{Landau \&  Lifshitz}{1987}]{LL1987} Landau, L.\ D.\ \&  Lifshitz, E.\ M.\ 1987, {\it Fluid Mechanics}, Vol. 6 (2nd ed.). Butterworth-Heinemann


\bibitem[\protect\citeauthoryear{Lodders}{2003}]{Lodders} Lodders K., 2003, ApJ, 591, 1220 


\bibitem[\protect\citeauthoryear{Markevitch \& Vikhlinin}{2007}]{2007PhR...443....1M} Markevitch M., Vikhlinin A., 2007, PhR, 443, 1 

\bibitem[\protect\citeauthoryear{Moekel}{1949}]{Moekel1949} Moekel W. E. 1949, Approximate Method for Predicting Form and Location of Detached Shock Waves Ahead of Plane or Axially Symmetric Bodies, NACA Technical Note 1921

\bibitem[Navarro et al.(1997)]{Navarro1997} Navarro, J.~F., Frenk, C.~S., \& White, S.~D.~M.\ 1997, ApJ, 490, 493
 
%The NGC 4839 group falling into the Coma cluster observed by XMM-Newton
\bibitem[\protect\citeauthoryear{Neumann et al.}{2001}]{Neumann.et.al.2001} Neumann D.~M., et al., 2001, A\&A, 365, L74 
\bibitem[\protect\citeauthoryear{Neumann et al.}{2003}]{Neumann.et.al.2003} Neumann D.~M., Lumb D.~H., Pratt G.~W., Briel U.~G., 2003, A\&A, 400, 811 


%First X-ray evidence for a shock at the Coma relic
\bibitem[\protect\citeauthoryear{Ogrean \& Br{\"u}ggen}{2013}]{Ogrean.Bruggen.2013} Ogrean G.~A., Br{\"u}ggen M., 2013, MNRAS, 433, 1701 

\bibitem[\protect\citeauthoryear{Planck Collaboration et al.}{2013}]{Planck} Planck Collaboration, et al., 2013, A\&A, 554, A140

\bibitem[\protect\citeauthoryear{Okabe et al.}{2014}]{2014PASJ...66...99O} Okabe N., et al., 2014, PASJ, 66, 99

\bibitem[\protect\citeauthoryear{Roediger et al.}{2015a}]{2015ApJ...806..103R} Roediger E., et al., 2015, ApJ, 806, 103 

\bibitem[\protect\citeauthoryear{Roediger et al.}{2015b}]{2015ApJ...806..104R} Roediger E., et al., 2015, ApJ, 806, 104 
  
%Suzaku Observations of Subhalos in the Coma Cluster
\bibitem[\protect\citeauthoryear{Sasaki et al.}{2015}]{Sasaki.et.al.2015} Sasaki T., Matsushita K., Sato K., Okabe N., 2015, ApJ, 806, 123
  
\bibitem[\protect\citeauthoryear{Sasaki et al.}{2016}]{Sasaki.et.al.2016} Sasaki T., Matsushita K., Sato K., Okabe N., 2016, PASJ, 68, 85 

%abundance of 0.3-0.4 solar
\bibitem[\protect\citeauthoryear{Sato et al.}{2011}]{Sato.et.al.2011} Sato T., Matsushita K., Ota N., Sato K., Nakazawa K., Sarazin C.~L., 2011, PASJ, 63, S991 

\bibitem[\protect\citeauthoryear{Sheardown et al.}{2019}]{Sheardown2019} Sheardown A., et al., 2019, arXiv, arXiv:1903.00482 

\bibitem[\protect\citeauthoryear{Sheardown et al.}{2018}]{2018ApJ...865..118S} Sheardown A., et al., 2018, ApJ, 865, 118 


%Thermodynamics of the Coma Cluster Outskirts
\bibitem[\protect\citeauthoryear{Simionescu et al.}{2013}]{Simionescu.et.al.2013} Simionescu A., et al., 2013, ApJ, 775, 4 


  
\bibitem[\protect\citeauthoryear{Smith et al.}{2001}]{APEC} Smith R.~K., Brickhouse N.~S., Liedahl D.~A., Raymond J.~C., 2001, ApJ, 556, L91 


\bibitem[Springel et al.(2001)]{Springel2001} Springel, V., Yoshida, N., \& White, S.~D.~M.\ 2001, New Astron., 6, 79


\bibitem[\protect\citeauthoryear{Springel, Frenk, \& White}{2006}]{2006Natur.440.1137S} Springel V., Frenk C.~S., White S.~D.~M., 2006, Natur, 440, 1137 

\bibitem[\protect\citeauthoryear{Su et al.}{2017}]{2017ApJ...835...19S} Su Y., et al., 2017, ApJ, 835, 19 


\bibitem[\protect\citeauthoryear{Sun et al.}{2006}]{2006ApJ...637L..81S} Sun M., Jones C., Forman W., Nulsen P.~E.~J., Donahue M., Voit G.~M., 2006, ApJ, 637, L81 

\bibitem[\protect\citeauthoryear{Sun et al.}{2009}]{2009ApJ...693.1142S} Sun M., Voit G.~M., Donahue M., Jones C., Forman W., Vikhlinin A., 2009, ApJ, 693, 1142   

\bibitem[\protect\citeauthoryear{Sun et al.}{2010}]{2010ApJ...708..946S} Sun M., Donahue M., Roediger E., Nulsen P.~E.~J., Voit G.~M., Sarazin C., Forman W., Jones C., 2010, ApJ, 708, 946 

\bibitem[Williamson(1996)]{Williamson1996} Williamson, C.~H.~K.\ 1996, Annual Review of Fluid Mechanics, 28, 477


\bibitem[\protect\citeauthoryear{van Dyke}{1982}]{1982aafm.book.....V} van Dyke M., 1982, aafm.book

\bibitem[\protect\citeauthoryear{van Weeren et al.}{2009}]{2009A&A...505..991V} van Weeren R.~J., R{\"o}ttgering H.~J.~A., Br{\"u}ggen M., Cohen A., 2009, A\&A, 505, 991 


\bibitem[\protect\citeauthoryear{van Weeren et al.}{2010}]{2010Sci...330..347V} van Weeren R.~J., R{\"o}ttgering H.~J.~A., Br{\"u}ggen M., Hoeft M., 2010, Sci, 330, 347 

\bibitem[\protect\citeauthoryear{Verigin et al.}{2003}]{Verigin2003} Verigin M., et al., 2003, JGRA, 108, 1323 



\bibitem[\protect\citeauthoryear{Vikhlinin, Forman, \& Jones}{1997}]{1997ApJ...474L...7V} Vikhlinin A., Forman W., Jones C., 1997, ApJ, 474, L7 



\bibitem[\protect\citeauthoryear{Vikhlinin et al.}{2005}]{Vikhlinin.et.al.2005} Vikhlinin A., Markevitch M., Murray S.~S., Jones C., Forman W., Van Speybroeck L., 2005, ApJ, 628, 655 

\bibitem[\protect\citeauthoryear{Vikhlinin et al.}{2006}]{2006ApJ...640..691V} Vikhlinin A., Kravtsov A., Forman W., Jones C., Markevitch M., Murray S.~S., Van Speybroeck L., 2006, ApJ, 640, 691   

\bibitem[\protect\citeauthoryear{Vikhlinin et al.}{2014}]{2014PhyU...57..317V} Vikhlinin A.~A., Kravtsov A.~V., Markevich M.~L., Sunyaev R.~A., Churazov E.~M., 2014, PhyU, 57, 317-341

 
\bibitem[Zhang et al.(2014)]{Zhang2014} Zhang, C., Yu, Q., \& Lu, Y.\ 2014, ApJ, 796, 138

\bibitem[Zhang et al.(2015)]{Zhang2015} Zhang, C., Yu, Q., \& Lu, Y.\ 2015, ApJ, 813, 129 

\bibitem[\protect\citeauthoryear{Zhang et al.}{2019}]{2018arXiv180802885Z} Zhang C., Churazov E., Forman W.~R., Jones C., 2019, MNRAS, 482, 20

\end{thebibliography}
\end{document}